\documentclass[aps,prd,superscriptaddress,reprint,nofootinbib]{revtex4-1}
\usepackage{graphicx}
\usepackage[table, dvipsnames]{xcolor}
\usepackage{amsmath}
\usepackage{amssymb}
\usepackage{siunitx}
\usepackage{orcidlink}
\usepackage[acronym]{glossaries}
\usepackage{hyperref}
\usepackage{cleveref}
\usepackage{ulem}
\usepackage{float}
\usepackage{amsfonts}

\usepackage{todonotes}

\usepackage{hyperref}
\hypersetup{
    colorlinks=true,
    linkcolor=blue,
    filecolor=magenta,      
    urlcolor=cyan,
    pdftitle={Overleaf Example},
    pdfpagemode=FullScreen,
    }

\begin{document}

\title{A pipeline for searching and fitting instrumental glitches in LISA data}

\author{Martina Muratore$\,$\orcidlink{0000-0002-9630-5698}}\email{contact: martina.muratore@aei.mpg.de}
\affiliation{\addressii}
\author{Jonathan Gair$\,$\orcidlink{0000-0002-1671-3668}}
\affiliation{\addressii}
\author{Olaf Hartwig$\,$\orcidlink{0000-0003-2670-3815}}
\affiliation{\addressaeihan}\affiliation{\addressunihan}
\author{Michael L. Katz$\,$\orcidlink{0000-0002-7605-5767}}
\affiliation{\addres}
\author{Alexandre Toubiana$\,$\orcidlink{0000-0002-2685-1538}}
\affiliation{\addressio} \affiliation{\addrese}
\newcommand{\at}[1]{\textcolor{red}{\bf [AT: #1]}}

\def\addressii{Max Planck Institute for Gravitational Physics (Albert Einstein Institute), D-14476 Potsdam, Germany}
\def\addressaeihan{Max-Planck-Institut für Gravitationsphysik (Albert-Einstein-Institut), Callinstraße 38, 30167 Hannover, Germany}
\def\addressunihan{Leibniz Universität Hannover, Institut für Gravitationsphysik, Callinstra\ss e 38, 30167 Hannover, Germany}
\def\addressio{Dipartimento di Fisica “G. Occhialini”, Universitá degli Studi di Milano-Bicocca, Piazza della Scienza 3, 20126 Milano, Italy}

\def\addres{NASA Marshall Space Flight Center, Huntsville, Alabama 35811, USA}

\def\addrese{INFN, Sezione di Milano-Bicocca, Piazza della Scienza 3, 20126 Milano, Italy}

\date{\today}

\begin{abstract}

\noindent Instrumental artefacts, such as glitches, can significantly compromise the scientific output of LISA. Our methodology employs advanced Bayesian techniques, including Reversible Jump Markov Chain Monte Carlo and parallel tempering to find and characterize glitches and astrophysical signals. The robustness of the pipeline is demonstrated through its ability to simultaneously handle diverse glitch morphologies and it is validated with a ``Spritz''-type data set from the LISA Data Challenge. Our approach enables accurate inference on Massive Black Hole Binaries, while simultaneously characterizing both instrumental artefacts and noise. These results present a significant development in strategies for differentiating between instrumental noise and astrophysical signals, which will ultimately improve the accuracy and reliability of source population analyses with LISA.
\end{abstract}

\maketitle
\section{Introduction}
\noindent The space-based gravitational wave detector, the Laser Interferometer Space Antenna (LISA) mission~\cite{amaroseoane2017laserinterferometerspaceantenna,colpi2024lisadefinitionstudyreport} was adopted as a mission by the European Space Agency in January 2024 and is scheduled for launch in 2035. One of its main objectives is the first direct observation of coalescing massive black holes binaries (MBHBs) with gravitational waves (GWs). Such observations will transform our understanding of the astrophysical processes leading to the formation of massive black holes and driving their coevolution with their host galaxies~\cite{Gair:2010bx,Sesana:2010wy,Klein:2015hvg,Dayal:2018gwg,Barausse:2020mdt,Toubiana:2021iuw,Chen:2022sae,Fang:2022cso,Spadaro:2024tve,Langen:2024ygz,Toubiana:2024bil}. Moreover, these observations will reach signal-to-noise ratios (SNRs) of thousands, allowing for exquisite tests of general relativity~\cite{Berti:2005ys,Berti:2016lat,Chamberlain:2017fjl,Barausse:2020rsu,Maggio:2020jml,Bhagwat:2021kwv,Corman:2021avn,Toubiana:2023cwr,Pitte:2024zbi}. Finally, the possible observation of electromagnetic counterparts~\cite{Mangiagli:2022niy,Dong-Paez:2023qlf,Izquierdo-Villalba:2024bhc} would allow MBHBs to be used as ``bright sirens'' for the measurement of cosmological parameters, not only the Hubble constant, but possibly also the equation of state of dark energy, thanks to the wide redshift range of these observations~\cite{Caprini:2016qxs,Tamanini:2016zlh,Mangiagli:2023ize}. To achieve these science objectives, it is crucial to develop a robust data analysis pipeline for LISA.  \\

\noindent  Unlike current ground-based GW detectors such as LIGO~\cite{2015CQGra..32g4001L}, VIRGO~\cite{VIRGO:2014yos} and KAGRA~\cite{Somiya:2011np,Aso:2013eba}. 
LISA will be a signal-dominated detector with multiple GW signals overlapping in time and frequency. Indeed, in addition to MBHBs, signals from $\sim 10^7$ Galactic binaries are expected to be present in the data stream, of which $\sim10^4$ will be individually resolved, and the rest will form a stochastic noise that will dominate over the instrumental noise around 1 mHz~\cite{Nelemans:2001hp,Ruiter:2007xx,Korol:2017qcx,Lamberts:2019nyk,Korol:2021pun,Toubiana:2024qxc}. The extreme-mass-ratio inspirals (EMRIs) of stellar-origin compact objects into massive black holes in the centres of galaxies will also generate signals in the LISA data that could last for the whole mission duration. The rate of EMRI events is very uncertain, ranging from a few to tens of thousands~\cite{Gair:2010yu,Amaro-Seoane:2012lgq,Babak:2017tow,Pan:2021ksp,Seoane:2024nus}, and at the upper end of this range EMRIs could perhaps even form a stochastic foreground~\cite{Bonetti:2020jku,Pozzoli:2023kxy}. Finally, the loudest and most massive of the stellar-origin black hole binary population being observed by ground-based detectors are also expected to be observable during their inspiral phase~\cite{Sesana:2016ljz,Wong:2018uwb,Gerosa:2019dbe,Toubiana:2020cqv,Buscicchio:2021dph,Toubiana:2022vpp,Buscicchio:2024asl} and a variety of cosmological backgrounds could significantly contribute to the total noise (see~\cite{LISACosmologyWorkingGroup:2022jok} and references therein). Moreover, we will not have a direct and independent measurement of the instrument noise \cite{PhysRevD.107.082004}, meaning that we will have to estimate its properties simultaneously with the parameters of the astrophysical sources. The simultaneous characterization of all the sources of multiple different types that are expected to be present in the LISA data, along with instrumental properties, is usually referred to as the LISA Global Fit~\cite{Cornish:2005qw,Littenberg:2023xpl,katz2024efficientgpuacceleratedmultisourceglobal,Strub:2024kbe,Deng:2025wgk}. \\
 
\noindent When designing a methodology to characterise instrumental noise, it is crucial to avoid systematic errors arising from noise misinterpretation or mismodeling. The LISA precursor mission, LISA Pathfinder (LPF) \cite{PhysRevD.110.042004}, which shares technology with LISA, was affected by spurious signals of unknown origin, referred to as glitches. Since various types of glitches were detected in LPF \cite{Armano_2022}, it is reasonable to expect that, unless they are fully understood and mitigated through hardware requirements, similar glitches may appear in LISA.
The LISA hardware, particularly the Gravitational Reference Sensors (GRS), will be largely the same as in LPF, with only minor modifications to accommodate integration into the Movable Optical Sub-assembly (MOSA) and the LISA spacecraft~\cite{Muratore:2021rwq}. Therefore, we anticipate encountering glitches similar to those seen in LPF—especially force impulses—which could impact LISA's sensitivity.\\

\noindent Glitches are also commonly seen in the data collected by ground-based gravitational wave detectors~\cite{Zevin:2016qwy}. These glitches show a complex range of morphologies, in contrast to the glitches observed in LPF, which were all of a much simpler form~\cite{PhysRevD.106.062001}. These differences most likely arise from the difference in the technology of the instruments and in their different operating environments. However, given the differences in the mission environment and duration from LISA and LPF, we should remain open to the possibility that LISA may encounter previously unobserved or more complex types of glitches. Regarding signals observed in ground-based detectors, these are typically very short in duration compared to their typical separation in time. For this reason, segments of data flagged as having glitches are normally excluded from scientific analysis. Some events, for example, the first binary neutron star merger, GW170817~\cite{LIGOScientific:2017vwq}, were clearly identified as being astrophysical in origin, due to good data quality in one of the instruments, while having overlapping glitches in one or more of the other detectors. In those cases, the analysis either gated out the range of times when the glitch was present or fitted out the glitch using a flexible model~\cite{Pankow:2018qpo}. In the case of LISA, signal durations will be typically much longer than the average time between glitches, and so only the latter strategies will be available to us.
\\

\noindent In this paper, we present a possible solution to address these types of artefacts—particularly force glitches acting on the test masses—within the framework of the Global Fit development for LISA. Our analysis is performed by using the open-source sampler \texttt{Eryn}, taking advantage of its reversible jump (RJ) capabilities \cite{Karnesis_2023,michael_katz_2023_7791640}. We demonstrate the feasibility of performing both searches and parameter estimation to fit multiple glitches of different shapes, morphology and SNR along with instrumental noise and an astrophysical source, in this case, an MBHB. \\

\noindent With this methodology, we aim to contribute to the ongoing efforts to develop robust data analysis techniques for LISA, ensuring its capacity to distinguish between astrophysical signals and instrumental artefacts. We therefore show the following  achievements:

\begin{enumerate}
\item Effective search, parameter estimation and fit of multiple glitches while recovering the LISA instrumental noise.  
\item Methodology to distinguish between a glitch and an MBHB signal. 
\item Preservation of parameter estimation accuracy for an MBHB signal in noisy data in the presence of one or multiple glitches, injected at various times—either before, after, or during the time of maximum overlap between the MBHB signal and the glitch.
\end{enumerate}
This paper is organised as follows: in section \ref{sec: glitch_lpf} we describe the LPF main results in terms of measured and catalogued artefacts; in section 
\ref{models} we introduce the models used in this paper for the instrumental noise, glitches, and the MBHB signal; in section \ref{sec:likelihood-formulation}, we provide details on the
Bayesian formalism used, including the likelihood and the sampling methodology; in section \ref{sec:data_analysis}, we present an example where we generate our own data containing only glitches and instrumental noise. This demonstrates the algorithm's ability to identify and fit multiple artefacts with different SNRs at the same time, while distinguishing them from stationary instrumental noise. Then, in section \ref{sec:estimation_part}, we analyse the impact of glitches with different SNRs on MBHB parameter estimation when left unmodeled and assess the recovery of both glitch and MBHB parameters when performing a joint fit. Finally, in section \ref{sec:spritz_data}, we apply our methodology to a ``\textit{Light}-Spritz'' Data Challenge \cite{LISA_LDC_Challenge2b}. We address it as ``\textit{Light}-Spritz'' because in this analysis
we ignore the non-stationarities caused by gaps and Galactic binaries present in the dataset, as these are subjects of a follow-up study. Additionally, we replace the MBHB signal included in this dataset with one generated using the PhenomHM \cite{London:2017bcn} waveform model from the BBHx software package~\cite{michael_katz_2023_7705496}, in order to eliminate systematics arising from differences between the simulated signal and the model used for the fit. We conclude the paper in section~\ref{sec:final} with a summary of our main results and perspectives for future work.

\section{Glitches as instrumental artefacts \label{sec: glitch_lpf}}
\noindent LPF successfully demonstrated the feasibility of maintaining free-falling test masses with residual acceleration noise below the stringent requirements for LISA ~\cite{PhysRevLett.120.061101}. Operating from 2016 to 2017, LPF achieved unprecedented noise levels; however, it also showed that measured(-experimental) noise could deviate quite significantly with respect to the derived noise models. In particular, the measured test-mass (TM) acceleration noise at frequencies below $10^{-4}$ is still not fully understood~\cite{PhysRevLett.120.061101,Armano_2024}. Also, transient acceleration glitches were observed during the mission \cite{Sala:2023hpr,Castelli:2020zro}. They spanned a wide amplitude range - transferring impulses from femto-Newton seconds (fN·s) to nano-Newton seconds (nN·s) to the test masses - and exhibited diverse durations, from seconds to hours. These glitches fell into two main categories: rapid transients in the interferometric readout and longer-lasting forces acting on the test masses. Glitches in the second category arose from spurious impulsive forces of unknown origin acting on the test masses, which occurred approximately once per day, following a Poisson distribution. These impulses exhibited a distinctive time evolution, with a peak followed by an exponential decay over several hours \cite{Armano_2022}. A detailed analysis of these phenomena showed that the observed signals could be accurately described using a fitted model \cite{Sala:2023hpr}, with minimal residual deviations. The strength of these impulsive events can be quantified through their SNR, which measures the detectability of a signal and will be formally defined later in the text in section~\ref{signal_to_noise}. \\
\noindent Out of the 432 events detected, 98 occurred during \textit{ordinary runs}, characterised by standard operational conditions (22 or 11 degrees Celsius). The SNR distribution of glitches that occurred during ordinary runs peaked around moderate values of 10. In contrast, more glitches (334) were observed during the \textit{cold runs} performed in May 2017, during which the instrument was operated at about 0 degrees. All glitches observed during the cold runs were short and intense, and 
had higher SNR values (see Figure 7.12 in \cite{Sala:2023hpr}). We refer the reader to \cite{Sala:2023hpr} for a more detailed discussion of the possible origin of this phenomenon. \\ 

\noindent In this paper, we focus primarily on glitches with morphology similar to those observed in the ordinary runs since LISA will operate at standard operational conditions. These exhibit a well-defined and predictable pattern. Indeed, the analysis of LPF glitches leveraged phenomenological modelling, as single or double exponentials with a certain decay time and amplitude \cite{Sala:2023hpr}. Recently, \cite{Baghi:2021tfd} has shown that shapelet functions could be used to fit LPF glitches and therefore, characterise their properties. These models, even if not directly connected to physical models as the ones derived in \cite{Sala:2023hpr,Castelli:2020zro}, laid the groundwork for future efforts to mitigate noise artefacts in LISA that are agnostic to the physical properties of the signals, such as the glitch decay time and impulse(-amplitude), which determine their shape. \\

\noindent To conclude this section, we note that one of the key challenges in GW data analysis is to develop methods that can robustly separate astrophysical signals from instrumental artefacts. To address this, we first need to derive a fiducial glitch model to use in our analysis. Therefore, in the next section, we present the shapelet function used to model the glitches and we relate it to the physical glitch model used to fit the observed LPF glitches. We also describe the waveform model used to describe the MBHB signal and the Time-Delay Interferometry (TDI) technique, which is applied to the raw data as a pre-processing step to reduce laser frequency noise before parameter estimation can be performed. 

\section{Time delay interferometry, glitches, Massive black hole binaries and noise models \label{models}}
\subsection{Time Delay Interferometry}
\noindent TDI is used in LISA to suppress laser frequency noise, which would otherwise exceed the GW signal by eight orders of magnitude, due to the inequality of the interferometer arms. The LISA mission comprises three identical spacecraft arranged in a triangular configuration, separated by 2.5 million kilometres and linked by six active laser connections. While the LISA satellite formation is designed to remain roughly equilateral, gravitational and orbital dynamics cause the arm lengths to fluctuate by approximately $1.5\%$ annually, with relative velocity drifts between spacecraft reaching up to 10 m/s \cite{amaroseoane2017laserinterferometerspaceantenna,colpi2024lisadefinitionstudyreport}. These time-varying and unequal arm lengths introduce laser frequency noise into the measurement channels. TDI achieves laser frequency noise reduction by forming specific linear combinations of time-delayed inter-spacecraft phase measurements \cite{PhysRevD.70.081101,Armstrong_1999}. Many combinations can be constructed \cite{Muratore:2020mdf}; however, for this work, we use the three noise-orthogonal channels $A$, $E$, and $T$ \cite{Prince:2002hp}, which are constructed from the Michelson TDI combinations $X$, $Y$, and $Z$, as these are the data streams available in ‘Spritz’.\\

\noindent Generally, when we refer to first-generation TDI \cite{PhysRevD.70.081101}, this means measurement combinations that are constructed to cancel laser frequency noise for a static, but unequal arm length, constellation. In contrast, for second-generation TDI, time-varying and unequal arm lengths are taken into account \cite{PhysRevD.68.061303,PhysRevD.72.042003}. In this paper, while we consider both generations, we assume in both cases equal and non-varying arm lengths. Specifically, the first-generation TDI is used for generating and analysing frequency data and modelling the noise and glitches in section \ref{sec:data_analysis}. Whereas, the second-generation TDI is used here for modelling the noise, signal, and glitches in the analysis of the ``\textit{Light-}Spritz'' Data Challenge \cite{LISA_LDC_Challenge2b} in section \ref{sec:spritz_data}.\\

\noindent Note that by assuming equal arm lengths, uncorrelated noise, and identical optical metrology system (OMS) and TM acceleration noise terms, the combinations $X, Y$, and $Z$ can be further transformed into three noise-orthogonal channels $A, E$, and $T$. This orthogonality simplifies the likelihood computation - in the frequency domain - by not having cross-correlation terms between channels. We will further discuss in the conclusion (section \ref{sec:final}) the effect if these assumptions are dropped. \\

\noindent The channels $A$, $E$ and $T$ \cite{PhysRevD.105.062006,PhysRevD.66.122002} that will be used here are:
\begin{align}
{\rm A} & = \frac{{\rm Z} - {\rm X}}{\sqrt{2}}\;, \hspace{3pt}  {\rm E} = \frac{{\rm X} - 2 {\rm Y} + {\rm Z}}{\sqrt{6}} \;, \hspace{3pt}
 {\rm T} & = \frac{{\rm X} +  {\rm Y} + {\rm Z}}{\sqrt{3}}.
\end{align}
In the case of first generation TDI, the channel $X$ is defined as:
\begin{align}\label{eq:tdi1-definition}
{\rm X_1}  &= (1-D_{12}D_{21} ) \times (\eta_{13} + D_{13} \eta_{31}) \nonumber
\\ & \qquad - (1 - D_{13}D_{31}) \times (\eta_{12} +  D_{12} \eta_{21}) \; ,
\end{align}
and the same variable in second generation is defined as: 
\begin{align}
{\rm X_2}  &=  (1-D_{12}D_{21} - D_{12}D_{21}D_{13}D_{31}
\nonumber
\\ &
\qquad \quad +D_{13}D_{31}D_{12}D_{21}D_{12}D_{21})\times (\eta_{13} + D_{13} \eta_{31}) \nonumber
\\ &\qquad - (1-D_{13}D_{31} - D_{13}D_{31}D_{12}D_{21}
\nonumber
\\ &
\qquad \quad +D_{12}D_{21}D_{13}D_{31}D_{13}D_{31})\times (\eta_{12} + D_{12} \eta_{21}) \nonumber
\\ & .\label{eq:tdi2-definition}
\end{align}
\noindent The $Y$ and $Z$ variables differ from $X$ by cyclic permutations of the three satellites. The operator $D_{pq}$ is a delay operator, corresponding to a constant time shift by $L_{pq}$ and thus in frequency to multiplication by $\mathcal{F}\{D_{pq}\} = e^{-i 2 \pi f L_{pq}}$ with $L_{pq}$ representing the light travel time for a beam received at spacecraft $p$ and emitted from spacecraft $q$, in seconds \cite{Heinzel_2011}. 
In this work we will assume the six arm-lengths to be equal and constant, and therefore $L_{pq} = L$.
The ${\eta}_{pq}$ are the so-called intermediary variables~\cite{Hartwig:2021dlc}. They represent a synthesised interferometric TM to TM measurement, here expressed as a fractional frequency shift between the interfering laser beams. The first index 
$p$ indicates the spacecraft where the measurement is performed at time $t$, and the second index $q$ indicates the distant spacecraft from which light was emitted at time $t - L_{pq}$. The intermediary variables ${\eta}_{pq}$ contain the sum of the contributions from GW signals, glitches and instrumental noises, and so the TDI transfer function applies linearly to these different components. 
\subsection{Shapelet model}
\noindent In this section, we describe how glitches are modeled and the LISA response to them. 
We model glitches using the exponential shapelet functions following \cite{Baghi:2021tfd} and \cite{Berge:2019nyt}. This is a family of exponentially damped functions that are eigen-wavefunctions of the normalised 1D-hydrogen atom. They can be defined in the time domain as follows:
\begin{equation}\label{eq:com_shap1}
\psi_n(t) = 2 \frac{t}{n} e^{-\frac{t}{n}}L^1_{n-1}\bigg (2 \frac{t}{n}\bigg)\Theta(t),
\end{equation}
with $L_n^\alpha$ denoting the generalized Laguerre polynomials and $\Theta(t)$ is the Heaviside step function. A glitch perturbation is then modeled by a linear combination of shapelets:
\begin{equation}\label{eq:glitch_shap1}
   g(t) = \sum_{j = 0}^{P} A_j \psi_{n_j} \bigg (\frac{t -\tau_j}{ \beta_j}\bigg),
\end{equation}
with $P+1$ being the number of shapelets. The parameters to be estimated for each shapelet component are the scale parameter $\beta_j$, which represents the characteristic damping time, the starting (or injection) time of the shapelet $\tau_j$, and the shapelet amplitude $A_j$.
Note that Eq.~\eqref{eq:glitch_shap1} is given in units of acceleration, but considering that the phasemeter in LISA outputs phase or frequency, we must convert it to the appropriate units. To achieve this, we utilise the generating function (see \cite{Wikipedia_LaguerrePolynomials} for reference and Appendix~\ref{shapelet_derivation}) and integrate Eq.~\eqref{eq:glitch_shap1} -after substituting Eq.~\eqref{eq:com_shap1}-  to derive the fractional frequency expression for a single exponential shapelet with $n_0 =1$ as: 
\begin{equation}\label{eq:integrated_dv}
\begin{split}
 &  \frac{ \Delta \nu_g}{\nu_0}(t,\tau_0, \beta_0, A_0)= \\
   &  \frac{2 A_0 \beta_0 }{c} 
       \bigg( 1 - \frac{e^{\frac{-t + \tau_0}{\beta_0}} (t + \beta_0 - \tau_0)}{\beta_0} \bigg) 
       \Theta \bigg(\frac{t - \tau_0}{\beta_0}\bigg),
\end{split}
\end{equation}

\noindent where we have considered that for our case $\alpha =1$ \cite{Baghi:2021tfd} and $\nu_0$ is the nominal laser frequency equal to 2.816e14 Hz. The shapelet model zero order depends thus on three parameters $\phi=\{\tau_0,A_0,\beta_0\}$.
We note that Eq.~\eqref{eq:integrated_dv} is the same as the one derived in~\cite{Baghi:2021tfd}. This function is similar to a step function starting from
a value of zero and increasing to:
\begin{equation}
 \lim_{t\rightarrow\infty}\frac{ \Delta \nu_g}{\nu_0}(t,\tau_0, \beta_0, A_0) =   \frac{2 A_0 \beta_0 }{c}.
 \end{equation}
The shapelet model with $j = 0$ thus reduces to the way glitches were modelled in LPF. Although, since the most significant impact of glitches during the LPF mission was an effective overestimation of the measured LPF acceleration noise, they had been fitted in terms of acceleration (Eq. 7.1 in \cite{Sala:2023hpr} and Appendix \ref{glitch_tdi}) using the following model: 
\begin{equation}
h_g(t) = \frac{\Delta v}{\tau^2}t' e^\frac{-t'}{\tau}\Theta(t'), \qquad \qquad t' = t - t_0
\end{equation}
which, integrated in time to get the fractional frequency change, gives:
\begin{align}\label{Eq:glitch_dv}
  \int_0^{t} & \frac{h_g(t')}{c} dt' \nonumber \\ & \hspace{0.5cm}= \frac{\Delta v}{c}
   \left(\frac{e^{\frac{t_0-t}{\tau }} (-t-\tau
   +t_0)}{\tau }+1\right) \Theta(t-t_0).
\end{align}
$\Delta v$ is the total transferred impulse per unit mass, $t_0$ is the injection time and $\tau$ is the decay time. Therefore, comparing the shapelet model (Eq. \ref{eq:integrated_dv}) with the glitch model (Eq. \ref{Eq:glitch_dv}), the shapelet amplitude $A_0$ can be related to the glitch impulse $\Delta v$ accordingly to $\Delta v  = 2 A_0 \beta_0 $, while $\beta_0$ equals the glitch $\tau$, and $\tau_0$ equals the glitch injection time $t_0$. These relationships will be important in section \ref{sec:data_analysis}, where we will use them to define the priors on the shapelet for fitting the glitches present in the data. Specifically, the prior for the shapelet amplitude will be chosen based on this relationship such that smaller amplitude values will be allowed compared to the nominal LPF glitches amplitude.\\
\begin{figure*}
\includegraphics[width=\textwidth]{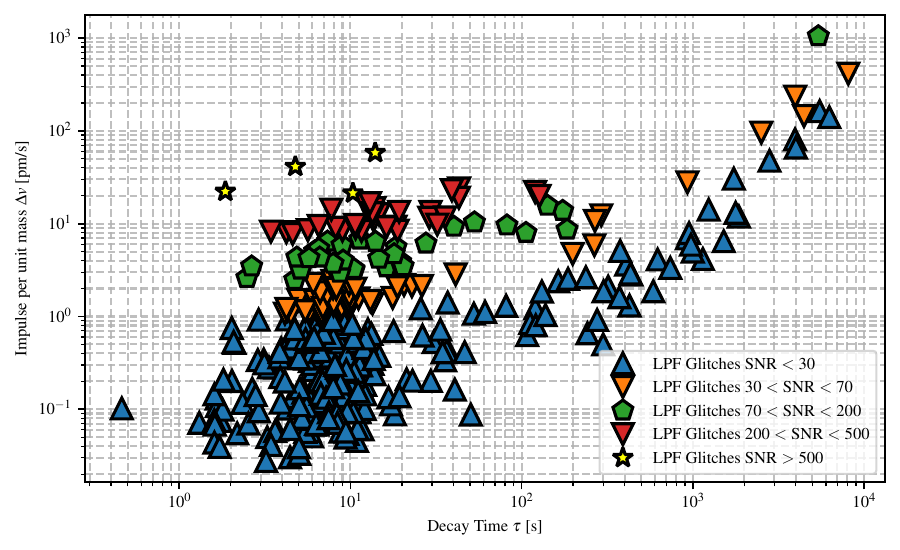}
 \caption{Impulse versus decay time of LPF glitches at different SNRs computed considering the LISA noise power spectral densities and the three TDI channels $A$, $E$ and $T$. \label{cap: threshold} }
\end{figure*} 

\begin{figure}[ht!]
\centering \includegraphics[width=0.95\columnwidth]{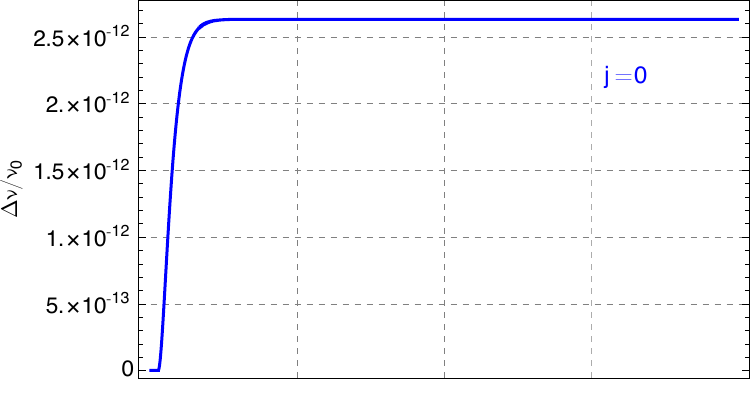}
\centering \includegraphics[width=\columnwidth]{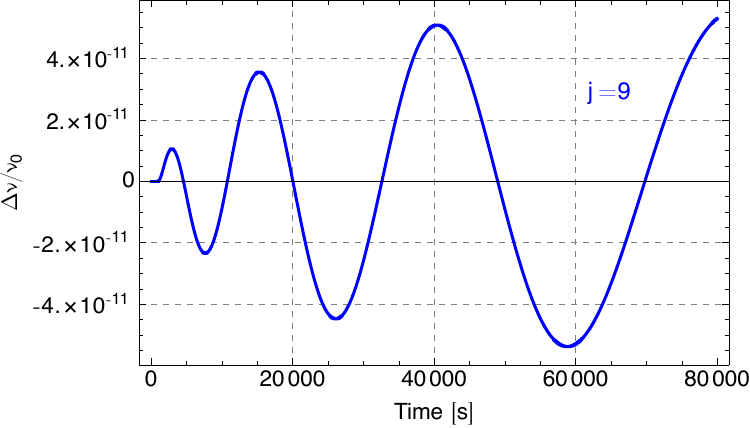}
\caption{\label{fig:shapelet}Shapelets with $n_0 = 1$ (upper plot) and $n_9 = 10$ (lower plot), where we have considered for both $A_0 = 5 \times 10^{-7} $, $\beta_0 =1000 $ and $\tau_0 = 1000 $. For the high order shapelet we assume all $A_j, \beta_j$ and $\tau_j$ to be equal to $A_0$ , $\beta_0$ and $\tau_0$, respectively.}
\end{figure}

\noindent Figure \ref{cap: threshold} shows the distribution of glitch parameters observed for LPF glitches. It is clear that relatively short glitches have the highest SNR. This is because short glitches have the majority of the power concentrated in the middle of the LISA frequency band, whereas long glitches concentrate the power at very low frequencies, where LISA is less sensitive. We note that SNRs for LISA are higher with respect to LPF (Fig.~\ref{cap: threshold} versus Fig.~8.8 in \cite{Sala:2023hpr}) for two reasons: the first is the different acceleration noise transfer functions between LISA and LPF, in particular, LISA shows a factor $4$ enhancement for the $X$ channels in the TM acceleration noise power spectral density (PSD) (Figure 8.8 in \cite{Sala:2023hpr}), the second is that we compute the SNR for LISA assuming the three TDI channels ($A$, $E$ and $T$) such that the SNR squared is doubled with respect to that in a single channel\footnote{The TDI channel $T$, or in general any channel null to GWs, being less sensitive to test-mass acceleration noise is also less sensitive to acceleration glitches. Therefore, their contribution to the SNR is negligible.}.\\
\noindent We also show in Figure~\ref{fig:shapelet}  two examples of shapelets with $j=0$ and $j=9$. 
\\

\noindent The shapelet expression (Eq. \ref{eq:integrated_dv}) can be written directly in the frequency domain as follows\footnote{We have tested that using the analytical frequency domain function is sufficiently equivalent to performing a numerical fast Fourier transform of the time domain data.}(see Appendix \ref{shapelet_derivation}):
\begin{equation}
\frac{\Delta \tilde{\nu}_g}{\nu_0}(f,\tau_0, \beta_0, A_0) =  \frac{i A_0 \beta_0 e^{-2 i f \pi \tau_0} }{
      c f \pi  ( i  - 2 f \pi \beta_0)^2}.\label{Eq.shap_freq}
\end{equation}
This expression agrees with the physical model used to fit glitches in LPF data but expressed in fractional frequency units (Eq. 7.3 in \cite{Sala:2023hpr} and Appendix \ref{glitch_tdi}). 

\subsection{TDI transfer function for glitches and shapelets in frequency domain\label{sec:tdi_tf}}

\noindent We can express the single link measurement $\tilde{\eta}_{pq}(\omega)$, in Eqs. \ref{eq:tdi1-definition},\ref{eq:tdi2-definition} as:
 \begin{equation}
	\tilde{\eta}_{pq}(f) = \Delta \tilde{\nu} _{g_{qp}}{(f)}e^{-2 i \pi f L}+\Delta \tilde{\nu}_{g_{pq}}{(f)},\label{eq:link}
\end{equation}
where the term $\Delta \tilde{\nu} _{g_{pq}}{(f)}$ indicates the contribution of a glitch (shapelet) acting on the test mass $pq$, and we have assumed constant and equal arm-lengths. Considering the expected glitch distributions from LPF, it is highly unlikely that multiple glitches happen simultaneously. We therefore assume that one of the two terms ($\Delta \tilde{\nu}_{g_{pq}}{(f)}$ or $\Delta \tilde{\nu} _{g_{qp}}{(f)}$) vanishes and that the other one contains a single glitch.\\

\noindent Considering now Eq. \ref{eq:tdi1-definition}, we can model the TDI applied to a glitch (shapelet) as a product between the glitch (shapelet) expression in the frequency domain (Eq. \ref{Eq.shap_freq}) and the transfer functions $\mathcal{T}_{X,Y,Z}^{g}(f)$ which take into account TDI, i.e.,  $\tilde{g}_{X,Y,Z}(f) = \mathcal{T}_{X,Y,Z}^{g}(f) {\Delta\tilde{\nu}_g}(f)/\nu_0$, where $\tilde{g}_{X,Y,Z}(f)$ is the signal of a glitch (shapelet) in the TDI frequency domain data (and we have omitted the constant glitch parameters in the argument of the expression). \\ 

\noindent In this work we will always apply glitches on $\rm TM_{12}$. Denoting by $\mathcal{F}$ the Fourier transform operator, and applying it to the delay operators gives $\mathcal{F}{D} = e^{-2 i \pi f L}$. The transfer function for the $X,Y$ and $Z$ first generation TDI variables of a glitch on $\rm TM_{12}$ is:
\begin{subequations}
\begin{align}
\mathcal{T}_{X_1}^{g}(f) & = -1 + e^{-8 i L \pi f} \\
\mathcal{T}_{Y_1}^{g}(f) & = 4 i e^{-4 i L \pi f} \sin(2 L \pi f) \\
\mathcal{T}_{Z_1}^{g}(f) & = 0.
\end{align}
\end{subequations}
The same type of derivation can be done for second-generation TDI using Eq. \ref{eq:tdi2-definition}:
\begin{subequations}
\begin{align}
\mathcal{T}_{X_2}^{g}(f) & = e^{-16 i L \pi f} (-1 + e^{8 i L \pi f})^2 \\ 
\mathcal{T}_{Y_2}^{g}(f) & = 8 e^{-8 i L \pi f} \sin(4 L \pi f) \sin(2 L  \pi f) \\
\mathcal{T}_{Z_2}^{g}(f) & = 0.
\end{align}
\end{subequations}
\noindent Note that, in the case of multiple glitches happening at the same time in different test masses, their effects sum up linearly. Expressions for the transfer functions for glitches occurring on the other test masses can be found in Appendix~\ref{glitch_tdi}.

\subsection{Waveform model for Massive Black Hole Binaries \label{waveform_model_BBHx}}

\noindent To model the full inspiral-merger-ringdown signal of a MBHB, we use the PhenomHM \cite{London:2017bcn} waveform model from the BBHx software package~\cite{michael_katz_2023_7705496}.
This is an analytic frequency-domain model for quasi-circular and spin-aligned binaries that includes harmonics beyond the dominant $(2,2)$ mode. The contribution of these harmonics increases with the asymmetry of the system (i.e., unequal mass ratios) and the magnitude of the spins. We do expect it to be possible to observe large mass ratio and highly spinning MBHBs with LISA~\cite{Toubiana:2021iuw,Barausse:2020mdt}, and not including higher harmonics could result in severe biases~\cite{Pitte:2023ltw}.   The GW signal of an MBHB, for the spin-aligned case, depends on 11 parameters $\theta = \{M_t, q, a_1, a_2, d_L, \varphi_{ref}, \iota, \lambda_d, \beta_d, \psi, t_{ref}\}$. They can be split as:
\begin{itemize}
    \item \textit{Intrinsic Parameters:}
    \begin{itemize}
    \item $m_1$ is the mass of the first black hole and $m_2$ of the second, with $m_2\leq m_1$. We express the masses as dimensionless multiples of the solar mass $\rm M_{\odot}$.
        \item \(a_1, a_2\): dimensionless spin magnitudes of the two black holes, aligned or anti-aligned with the orbital angular momentum.
    \end{itemize}
    \item \textit{Extrinsic Parameters:}
    \begin{itemize}
        \item $d_L$: luminosity distance to the source, scaling the amplitude of the GW signal.
        \item $\varphi_{ref}$: reference phase at the merger.
        \item \(\iota\): inclination, the angle between the orbital angular momentum and the line of sight, affecting the observed amplitude of higher-order modes.
        \item \(\lambda_d, \beta_d\): ecliptic longitude and latitude, describing the sky location of the source in the (LISA) detector reference frame.
        \item $\psi$: polarisation angle of the waveform, affecting the projection of the signal onto the detector.
        \item $t_{ref}$: reference time marking a characteristic point in the binary coalescence, such as the merger.
    \end{itemize}
\end{itemize}
\noindent All extrinsic parameters discussed here are defined relative to the LISA constellation reference frame. We also introduce the total mass $M_t = m_1 + m_2$ and the mass ratio $q = m_2/m_1 \leq 1$. \\
       
\noindent The frequency-domain, source-frame waveform is expressed as \cite{marsat2018fourierdomainmodulationsdelaysgravitationalwave,Marsat_2021}:  
\begin{equation}  
\tilde{h}_{lm}(f) = A_{lm}(f)e^{- i \phi_{lm}(f)},  
\end{equation}  
where \((l,m)\) represent the harmonic mode indices.
The TDI variables are obtained as 
\begin{equation}  
\tilde{h}_{X,Y,Z}(f) = \sum_{lm}\mathcal{T}^h_{X,Y,Z}(f,t_{lm}(f))\frac{\tilde{h}_{lm}(f)}{d_L}.  
\end{equation} 
where $\mathcal{T}_{X,Y,Z}^h(f, t_{lm}(f))$ are the frequency-domain transfer functions detailed in \cite{marsat2018fourierdomainmodulationsdelaysgravitationalwave, Marsat_2021}. The transfer functions depend on the extrinsic parameters and vary both in time and frequency due to the LISA constellation’s orbital motion. They are evaluated at time $t_{lm}(f)$, which is given by the stationary-phase approximation:  
\begin{equation}  
t_{lm}(f) = t_{ref} - \frac{1}{2 \pi} \frac{d \phi_{lm}(f)}{df}. 
\end{equation}  
\noindent The phase-meter output (LISA observables) will therefore be the sum of GW signals, $\tilde{h}_{X,Y,Z}(f)$, and glitches, $\tilde{g}_{X,Y,Z}(f)$, propagated through TDI.
\subsection{Noise models and definition of the signal to noise ratio \label{noise_model}}
\noindent In this work, we assume that primary noise sources such as laser frequency noise, spacecraft longitudinal jitters, clock noises and tilt-to-length couplings have been corrected and only leave negligible residuals in the post-processed TDI variables~\cite{PhysRevD.103.123027,Hartig_2022,PhysRevApplied.14.014030}. What remains are the so-called secondary noises. These remaining noises~\cite{PhysRevD.108.082004} fall into two broad categories: residual acceleration noise of each TM and the OMS noise terms limiting the TM position readout. Noises inside these categories can have various physical origins, different levels and different transfer functions into the final TDI variables~\cite{Quang_Nam_2023}. For the analysis that will follow in section \ref{sec:data_analysis}, we just consider two main components modelled as uncorrelated noise on each TM and uncorrelated noise in each inter-spacecraft optical link, with identical statistical properties for noises of the same kind. We will instead consider a slightly more elaborate noise model for analysing the ``\textit{Light-}Spritz'' data set, in section \ref{sec:spritz-noise}.\\

\noindent We characterise these noise sources by their PSD and a corresponding transfer function that describes their propagation in the TDI variables.  The TM acceleration noise PSD is:
\begin{equation}
\begin{split}
S_{\rm TM}(f) = (N_{\rm TM})^2& \left( 1 + \left(\frac{0.4 \times 10^{-3}}{f} \right)^2 \right) 
\\
\times & \left( 1 + \left(\frac{f}{8 \times 10^{-3}} \right)^4 \right),
\end{split}
\end{equation}
with $N_{\rm TM} = \SI{2.4}{\femto\meter\per\second\squared\per\sqrt\hertz}$. 

\noindent The PSD of the displacement noise from the OMS is:
\begin{equation}
S_{\rm OMS}(f) = (N_{\rm OMS})^2 \left(1 + \left(\frac{2.0 \times 10^{-3}}{f}\right)^4\right) ,
\end{equation}
with $N_{\rm OMS} = \SI{7.9}{\pico\meter\per\sqrt\hertz}$. This overall OMS noise level is based on the values provided in the LDC tools~\cite{lisa_data_challenge_working_group_2022_7332221}. \\

\noindent Note that the noise parameters, $\gamma = \{N_{\rm TM}, N_{\rm OMS}\}$, are expressed as equivalent acceleration noise in \si{\meter\squared\per\second^4 \per\hertz} or displacement noise in \si{\meter\squared\per\hertz}. For our analysis, we instead use units of fractional frequency deviations, to which we will convert them by multiplying the components with $C_{\mathrm{acc}}^{\mathrm{ffd}} = \left({1}/{(c 2 \pi f)}\right)^2$ or $C_{\mathrm{disp}}^{\mathrm{ffd}} = \left({2 \pi f}/{c}\right)^2$, respectively. Here, $c$ is the speed of light in \si{\meter\per\second}.\\

\noindent For the first generation TDI, the noise in the $A, E$ and $T$ channels is:
\begin{align}
S_{A}(f) =& S_{E}(f) = 8 \sin^2( 2\pi L f) \Big[
\nonumber \\
& 2C_{\mathrm{acc}}^{\mathrm{ffd}}  S_{\rm TM} (3 + 2\cos( 2\pi L f) + \cos( 4\pi L f))   
\\ 
& +C_{\mathrm{disp}}^{\mathrm{ffd}}  S_{\text{OMS}} (2 + \cos( 2\pi L f)) \Big], \nonumber
\end{align}
and:
\begin{align}
S_T(f) = & 16 C_{\mathrm{disp}}^{\mathrm{ffd}} 
 S_{\rm{OMS}} (1 - \cos( 2\pi L f)) \sin^2( 2\pi L f)  \nonumber \\ & +  128 C_{\mathrm{acc}}^{\mathrm{ffd}} S_{\rm{TM}} \sin^2( 2\pi L f) \sin^4\left( \pi L f \right).
\end{align}
Where, again, $L$ denotes the LISA arm length in seconds. 
\subsubsection{Definition of the signal to noise ratio \label{signal_to_noise}}
\noindent The SNRs of glitches and MBHB signals are computed based on the three TDI channels using the following formula (Eq. 119 in \cite{tinto} and \cite{babak2021lisa}):
\begin{align}
\rm SNR = & \Big[ 4 \Re \sum_{f}|\tilde{Q}(f)|^2 \Big( \frac{|\mathcal{\tilde{T}}_{\rm A}(f)|^2}{S_{\rm A}(f)} + \frac{|\mathcal{\tilde{T}}_{\rm E}(f)|^2}{S_{\rm E}(f)}   \notag \\
& \quad  + \frac{|\mathcal{\tilde{T}}_{\rm T}(f)|^2}{S_{\rm T}(f)} \Big) \, \Delta f \Big]^{1/2}.\label{eq:SNRdef}
\end{align}
where $\Delta f$ is the frequency resolution which is equal to $1/T_{\rm obs}=1/(N dt)$, with $T_{\rm obs}$ the observation time, $N$ the length of the time series and $dt$ is the inverse of the sampling rate.
Note that the above formula is valid in this case as we assume the three TDI channels to be uncorrelated.
Here $\mathcal{\tilde{T}}_{A/E/T}(f)$ are the transfer functions of a GW signal or a glitch into the combinations $A$, $E$, and $T$ and $\tilde{Q}(f)$ is the amplitude of the GW signal ($\tilde{h}_{lm}(f)/d_L$) or glitch ($\Delta \tilde{\nu}_g(f)/\nu_0 $) in Fourier frequency domain, and $S_{A/E/T}(f)$ are the spectral densities of the noises in the $A$, $E$, $T$ combinations respectively. All these quantities have been defined in the previous sections.

\section{Bayesian methodology and likelihood formulation}
\label{sec:likelihood-formulation}

\noindent The Bayesian framework provides a powerful approach to the problem we aim to solve by treating the detection and parameter estimation of GW signals, instrumental artefacts and noise as an inference problem. In this context, the observed data, $d(t)$, is modelled as the sum of contributions from the GW signal, $h(t, \theta)$, the glitch, $g(t, \phi)$, and the noise, $n(t,\gamma)$:
\begin{equation}
d(t,\theta,\phi, \gamma) = h(t, \theta) + g(t, \phi) + n(t, \gamma)\label{data_time},
\end{equation}
where $\theta,\phi$ and $\gamma$ represent the parameters of the GW signal, the glitch and the noise model, respectively. Denoting by $\mu=(\theta, \phi,\gamma)$ the set of all parameters, we write $p(\mu| d,M)$ the posterior distribution of the parameters. We note $M$ the assumed model, which includes $M_1$, our chosen shapelet model for the glitches, $M_2$, our model for the MBHB signal, and $M_3$, our model for the instrumental noise. The posterior distribution $p(\mu | d, M$) is computed using Bayes’ theorem \cite{SiviaSkilling2006,Karnesis_2023}:
\begin{equation}
  p(\mu | d, M) =
    \frac{p(d | \mu,M) \, p(\mu | M)}
    {p(d | M)}.  
\end{equation}
where:
\begin{itemize}
    \item $p(d | \mu, M )$ is the likelihood, quantifying the probability of observing data $d$ under the models $M_1$, $M_2$ and  $M_3$ with parameters $\theta, \phi$, and $\gamma$;
    \item $p(\mu | M)$ is the prior distribution for the GW, glitch and noise parameters under the models $M_1$, $M_2$ and $M_3$: $p(\theta | M_1)$, 
    $p(\phi| M_2)$ 
    and $p(\gamma| M_3)$;
    \item $p(d| M)$ is the evidence, which acts as a normalization constant for parameter estimation, but can also be used to compare different models $M$. 
\end{itemize}

\noindent For the problem at hand, we will use the Whittle likelihood, for which the log-likelihood in the frequency domain is:
\begin{align}
\log \mathcal{L} = -\frac{1}{2} \sum_{f} \Bigg[ & 4 \Delta f \frac{\left| \tilde{d}(f) - \tilde{h}(f, \theta) - \tilde{g}(f, \phi) \right|^2}{S(f, \gamma)} \nonumber  \\
& + \log S(f, \gamma) \Bigg], 
\end{align}
where $\tilde{d}(f), \tilde{h}(f, \theta)$ and $\tilde{g}(f, \phi)$ are the Fourier transforms of the same quantities appearing in Eq. \ref{data_time}; $S(f, \gamma)$ is the one-sided noise PSD, parameterized by $\gamma$.  \\

\noindent The posterior can be sampled using advanced Bayesian inference techniques such as Markov Chain Monte Carlo (MCMC) \cite{Metropolis1953,Hastings1970,Hitchcock2003,Gilks1995}. In this analysis, we wish to fit simultaneously an unknown number of glitches. Therefore, we use a Reversible Jump MCMC (RJ-MCMC)~\cite{Green1995} algorithm, a generalisation of the Metropolis–Hastings algorithm that allows us to vary the dimensionality of the parameter space while sampling over the parameters. We use the RJ-MCMC implementation of \texttt{Eryn}~\cite{Karnesis_2023}. Several walkers are run in parallel and exchange information periodically in order to increase the sampling efficiency. We also use parallel tempering~\cite{PhysRevLett.57.2607,doi:10.1143/JPSJ.65.1604,Vousden2015} to improve the exploration of multi-modalities in the posterior distribution. 
During the run, a new point $\mu'$ is proposed from a point $\mu$ according to a proposal function, $q(\mu ' |\mu)$, and is accepted with probability:
\begin{equation}
\alpha = \min\left( 1, \frac{p( \mu'|d,M) \, q(\mu|\mu')}{p( \mu|d,M) \, q(\mu ' | \mu)} \right).\label{eq:acep_ratio}
\end{equation}
In our analysis, we use different proposals for the parameters of the MBHB ($\theta$), of the instrumental noise ($\gamma$) and of the glitches ($\phi$). They are updated successively, using the Gibbs sampling implementation of \texttt{Eryn} \cite{Karnesis_2023}. In the \texttt{Eryn} `language', branches represent different types of models used to fit the data. Each branch contains leaves, which correspond to individual instances of each model. In our case, leaves represent individual glitches, massive black hole binaries, or noise components, which form different branches~\cite{Karnesis_2023}.

\subsection{Moves}\label{sec:moves}
In an RJ-MCMC sampler, there are two types of moves: 
\begin{itemize}
\item In-model moves: changes are made to the parameters, keeping the number of leaves in each branch fixed. 
\item Out-of-model moves: the dimensionality is varied by adding or removing leaves, in one or several branches. For instance, in our case, adding or removing glitches. 
 \end{itemize}
 In this work, only the number of glitches is varied, so that the latter type of moves applies only to their parameters. Detailed explanations of the various moves available in \texttt{Eryn}, as well as the most commonly used ones in the literature, can be found in~\cite{michael_katz_2023_7705496,Karnesis_2023}. Here, we focus on the in-model moves specifically developed for glitch searches, MBHB parameter estimation, and noise parameter estimation. For the out-of-model move for glitches, we propose new parameters from the prior distribution during the search phase. Instead, during the final parameter estimation phase, as explained in the upcoming section, we propose sampling from a Gaussian distribution, whose covariance matrix is constructed from an initial estimate of the parameters of the detected glitches during the search phase.

\subsubsection{\textbf{Glitch search and parameter estimation (in-model moves)}\label{glitch_in_model}}
\subsubsection*{Group stretch move}
\noindent In the stretch move, new samples are generated by leveraging the distribution of points in the ensemble, using the formula:
\begin{equation}
    Y = X_j + z (X_k - X_j), \label{eq:stretch_move}
\end{equation}
where $X_k$ is the current sample, $X_j$ is another point in the ensemble, $z$ is a scaling factor drawn from a predefined distribution, and $Y$ is the proposed new value. In RJ-MCMC, however, the variable dimensionality of models complicates the direct use of ensemble-based proposals. To deal with the variable dimensionality of models, the Group Stretch Move is used \cite{Karnesis_2023}, which builds upon the affine-invariant stretch proposal \cite{2010CAMCS...5...65G}. This group introduces `friends', a static subset of reference points that are defined after a first burn of the sampler and as soon as the group move is used. These friends simulate the posterior distribution for a given model and remain fixed throughout multiple iterations. For instance, in our case, we group the estimated glitches into `friend groups' according to their mean time value. The goal is to set $X_j$ properly for each posterior mode such that $X_j$ and $X_k$ are always found on the same mode of the posterior. This approach improves efficiency by ensuring that proposed points correspond to the posterior distribution of the currently sampled glitch, rather than a different one. \\ The stretch move is used 50$\%$ of the time, and in the implementation of this move, the set of `friends' associated with each glitch is tracked\footnote{To best use this move, we leverage the branch supplemental object that has the goal of tracking the friends' information associated with each glitch. This is particularly needed in the presence of RJ-MCMC and parallel tempering, where we have swaps in temperature or birth/death of new glitches}, to improve the efficiency of the RJ-MCMC to add and remove glitches~\cite{Karnesis_2023, michael_katz_2023_7791640}.

\subsubsection*{Selected Covariance Group Move}
\noindent The `Selected Covariance Group Move' replaces the standard Gaussian move. In the context of nested models, the posterior distribution may display distinct modes, corresponding to different glitches in the data. This complexity makes it challenging to efficiently tune covariance matrices for each mode. The core idea behind this move is to group glitches based on their mean injection time values and compute corresponding distributions for glitch parameters using \href{https://scikit-learn.org/stable/modules/mixture.html}{a Gaussian Mixture Model (GMM)}. This approach leverages covariance matrices specific to each glitch group. When glitches are clearly distinguishable, the group proposal becomes highly efficient, aided by the branch supplemental object in \texttt{Eryn}~\cite{michael_katz_2023_7705496} to maintain the association between glitches and their respective covariance matrices. Note that we leverage the ``fix friends'' function from \texttt{Eryn} when a proposed birth of a new glitch is accepted. This function assigns a covariance matrix (or friends) to the new glitch from the precomputed friends group.\\

\noindent The computation(-update) of the covariance matrices for this move is done as follows: 

\begin{enumerate}
\item Cluster specification: Every 500 iterations, we evaluate the fraction of samples for which the number of active glitch leaves is zero — i.e., samples where no glitch components are present. Physically, this corresponds to samples where the data is explained entirely by the noise model, without requiring any glitch contribution \cite{Karnesis_2023}. 
If this fraction is sufficiently bigger than zero, it indicates that a significant portion of the current samples includes one or more active glitch components — suggesting that the sampler interprets the data as containing non-Gaussian features such as glitches. In this case, a GMM is initialised using the posterior samples, with the number of mixture components set equal to the maximum number of glitch leaves allowed by the model. This enables the sampler to model complex posterior structures as a combination of multiple Gaussian components. Each component in the GMM is assigned a full-rank covariance matrix and is intended to represent a distinct glitch cluster, with the number of components constrained by the dimensionality of the glitch model.

\item Covariance Calculation: Based on the identified number of clusters, a GMM is fitted to the data. The GMM provides the mean and covariance for each cluster, capturing the local structure of the posterior around each glitch\footnote{In a GMM, each Gaussian component does not necessarily correspond to an independent cluster. If a posterior distribution is not very Gaussian in shape, more than one component may be needed to approximate it. Therefore, identifying a GMM component is not the same as identifying a true cluster represented by the posterior of a single glitch. However, for the problem analysed here, we have verified that the glitch posteriors are sufficiently Gaussian and well-separated, so our assumptions remain valid. Dividing a true cluster into more sub-clusters than are actually present in the data leads to proposal steps that are smaller than necessary. This can result in a high acceptance rate but a low effective sample size, meaning the sampler requires more steps to converge. To improve sampling efficiency, this proposal method should be refined in future work. One development we have already begun testing is using x-means instead of \href{https://scikit-learn.org/stable/modules/generated/sklearn.cluster.KMeans.html}{k-means} to automatically determine the optimal number of GMM components. 
}. These parameters are then utilised in the in-model move, enabling it to propose moves that are well-aligned with the posterior’s structure. This approach ensures the sampler's efficiency, particularly when the shape of the posterior varies significantly between glitches, warranting the use of distinct covariance matrices for each glitch. By iteratively estimating the covariance matrices from the samples via a GMM, and adapting the sampler to the posterior structure, this methodology establishes a robust framework suitable for glitch searches. Note that adaptation is applied for the first 50000 iterations -every 500 iterations- and then stopped to preserve detailed balance and ensure the Markovianity of the move. The covariance group move is used 50$\%$ of the time.
\end{enumerate}

\subsubsection{\textbf{Glitch search and parameter estimation (out-of model moves) \label{ref:rj_move_glitch}}}
\noindent During the parameter estimation phase, we incorporate a move proposal mechanism based on detected glitches. Specifically, $70\%$ of the time, the glitch parameters are drawn from a mixture of independent Gaussian distributions describing the shapelet parameter $\tau_0$, $\ln(\rm A_0)$ and $\beta_0$. Each parameter is sampled from a multivariate normal distribution with the corresponding mean and covariance of the previously detected shapelets. We propose from the prior distribution the other $30\%$ of the time. 
 
\subsubsection{\textbf{Noise parameter estimation (in-model moves)}\label{in_model_noise}}
  \noindent We consider both the stretch and Gaussian proposal moves \cite{michael_katz_2023_7705496}. Initially, the Gaussian move uses a diagonal covariance matrix with small entries (following a suggestion from the \href{https://mikekatz04.github.io/Eryn/html/tutorial/Eryn_tutorial.html#Add-multiple-branches}{Eryn tutorial}). Then, the covariance matrix is re-estimated using the noise samples every 500 iterations, and the adaptation is done for the first 50000 iterations to preserve detailed balance. The stretch move is used 85$\%$ of the time while the Gaussian move 15$\%$.
  
\subsubsection{\textbf{Massive black hole binaries parameter estimation (in-model moves)} \label{proposal_mbh}}
\noindent It is known that at low frequencies, MBHB posteriors have an eightfold sky mode degeneracy. This is reduced to a twofold latitudinal degeneracy at higher frequency. To ensure efficient sampling of the sky-mode posterior, following Ref.~\cite{PhysRevD.105.044055}, we therefore use a sky-mode hopping proposal in addition to the stretch proposal described above. The stretch proposal is used in approximately $80\%$ of proposals. We also add a Gaussian move proposal used $5\%$ of the time with a small covariance matrix. This matrix is then recursively estimated from the MBHB samples with the same cadence used to estimate the noise covariance matrix. We notice that this additional move helps the convergence of the algorithm in possible cases where $X_k$ and $X_j$ (Eq. \ref{eq:stretch_move}) are very close in value, as they will spread out slowly otherwise if we only use the stretch move.  Instead, the covariance Gaussian move will speed this up. \\

\noindent Regarding the sky-mode proposal, when sampling in the LISA reference frame \cite{Marsat_2021}, as we do here, the eight distinct sky modes consist of four longitudinal modes at:
\[
\{\lambda_d + (0, 1, 2, 3)\pi/2, \psi + (0, 1, 2, 3)\pi/2\}
\] 
and two latitudinal modes: 
\[
\{\pm \beta_d, \pm \cos \iota, \pm \cos \psi\}.
\]

\noindent Three varieties of this proposal are used: 
\begin{itemize}
  \item One is to move the walker to a mode drawn randomly with replacement from all eight sky modes. This is used for approximately $4\%$ of proposals.
  \item Proposals to change just the longitudinal or latitudinal mode are used approximately $3\%$ and $8\%$ of the time, respectively.
\end{itemize}

\noindent The relative percentages of the sky-mode hopping proposals reflect expectations about their relative importance~\cite{PhysRevD.105.044055}. 

\section{Testing the reversible jump with simulated data}\label{sec:data_analysis}
\noindent To test our pipeline, we generate mock data by injecting glitches—extracted from the LPF catalogue~\cite{PhysRevD.106.062001} and shown in Fig.~\ref{cap: threshold}—into a noise-only dataset. For this test, we do not include any astrophysical signals, as our goal is to investigate the ability of the RJ-MCMC to identify glitches of varying shapes and SNRs, and to distinguish them from noise in order to establish an approximate threshold below which such glitches are consistent with instrumental noise. In Section~\ref{sec:estimation_part}, we also quantify the impact of three fiducial glitches—with low, medium, and high SNRs of 21, 72, and 1544, respectively—on the parameter estimation of an MBHB. 
\subsection{Data generation}
\noindent We generate 30 days of LISA data where seven glitches are injected according to the amplitudes and damping times extracted from the LPF catalogue \cite{PhysRevD.106.062001}, and are randomly distributed over these 30 days. In LPF, the glitches were observed to have a Poisson distribution with about one glitch per test mass per day. We use a slightly lower rate here to simplify the computational burden as we do not believe this will affect our results qualitatively.
Moreover, we also allow for glitches injected very close in time to show the ability of the methodology to distinguish between such glitches. \\
\noindent Table \ref{tab:glitches} reports the parameters of the injected glitches as well as the corresponding number indicating the position of the glitch in the LPF catalogue (the run index is in accordance with LPF numbering). \\

\noindent We model the injected glitches using the single exponential model reported in section~\ref{models} Eq. \ref{Eq:glitch_dv}, but we use shapelets with $n_0=1$ to fit for them. 
 \begin{figure}[h]
    \centering
        \includegraphics[width=\columnwidth]{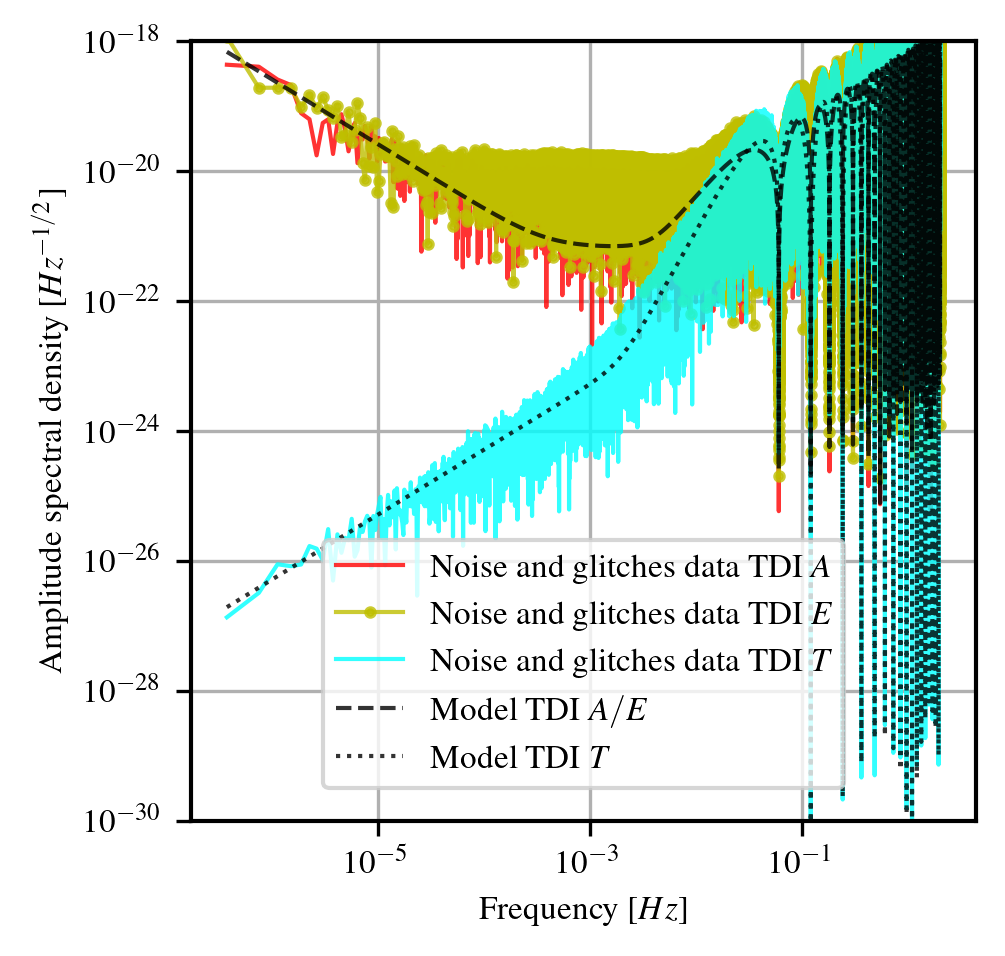}
 \caption{Amplitude spectral density of the TDI channels $A, E$ and $T$ when seven glitches are injected on $\rm {TM}_{12}$ at different times.\label{noise_with_glitches}}
 \end{figure}
We simulate LISA noise directly in the frequency domain following \cite{PhysRevD.109.042001}. Assuming that the real-time series $n(t)$ follows a stationary, zero-mean, Gaussian and ergodic random process, the Fourier transform of the noise, $\tilde n(f)$ is normally distributed with zero-mean and variance given by the one-sided power spectral density:
\begin{subequations}
\begin{align}
<\tilde{n}^{*}(f') \tilde{n}(f)> & = \frac{1}{2}S_n(f)\delta(f-f'), \label{eq:psd}\\
< \tilde{n}(f') \tilde{n}(f) > & = 0,
\end{align}   
\end{subequations}
where $<>$ denotes the expectation value over the data-generating process.  
For a finite grid of frequencies, defining $n_k = \tilde n(f_k)$, Eq.~\eqref{eq:psd} can be written as
\begin{align}\label{eqapp:variance_noise}
<\tilde{n}^{*}_k \tilde{n}_j> & = \frac{T}{2}S_n(f_k) \delta_{jk}.
\end{align}
Therefore the variance of the real and imaginary parts of the noise is given by (where the real and imaginary part are uncorrelated):
\begin{equation}\label{eqapp:variance_re_im_noise}
<\Re [\tilde{n}_k]^2>  = <\Im [\tilde{n}_k]^2> =  \frac{T}{4}S_n(f_k) = \sigma_k^2 \, .
\end{equation}
The noise in the Fourier domain is then generated as:
\begin{equation}
\tilde{n}_k   = \sigma_k \cdot \left( X_{\text{real}} + i X_{\text{imag}} \right),
\end{equation}
where $X_{\text{real}} \sim \mathcal{N}(0, 1)$ and $X_{\text{imag}} \sim \mathcal{N}(0, 1)$ are independent normally distributed random variables with mean 0 and variance 1. \\

\noindent For this example, the noise PSD is given by the sum of the TM acceleration and OMS noises which follow the models described in section \ref{noise_model}. The assumed value for the TM acceleration noise is 
$N_{\rm TM}^2 =  \left(2.4 \times 10^{-15}~\textrm{m}\,\textrm{s}^{-2}\textrm{Hz}^{-1/2}\right)^2 $ and for the  OMS noise $N_{\rm OMS}^2 = \left( 7.9 ~\textrm{pm}/\sqrt{\textrm{Hz}}\right)^2$.\\

\noindent Figure~\ref{noise_with_glitches} shows the Amplitude Spectral Density (ASD) in the TDI channels resulting from the injected glitches. 
The channels $A$ and $E$ show an excess of power with respect to the reference noise model in nominal conditions, shown in black. This excess of power is mild in the null channel $T$, as anticipated in \cite{Muratore:2021rwq}. The reason is that for null channels (acceleration-)glitches have a mild effect due to the OMS and TM acceleration noise transfer functions of these channels. In particular, the latter shows a suppression at low frequencies with respect to the OMS such that the OMS noise dominates for these channels at all frequencies \cite{PhysRevD.107.082004}. 
\begin{table}[ht!]
\centering
\begin{tabular}{@{}lcccc@{}}
\toprule
\textbf{Glitch} & \boldmath$t_0$ $[s]$ & \boldmath$\ln(\Delta v/[m s^{-2}])$ & \boldmath$\tau$ $[s]$ & \textbf{SNR} \\ 
\#3 & $2.97 \times 10^5$ & $-25.5$ & $184.47$  & $72$  \\ 
  \#118 & $7.50 \times 10^5$ & $-25.5$ & $4.23$  & $230$ \\ 
  \#90& $1.25 \times 10^6$ & $-24.4$ & $43.21$  & $479$ \\ 
 \#72& $1.75 \times 10^6$ & $-24.5$ & $1.86$  & $637$ \\ 
  \#179& $2.20 \times 10^6$ & $-23.9$ & $4.77$  & $1172$ \\ 
  \#248 & $2.25 \times 10^6$ & $-23.6$ & $13.96$  & $1544$ \\ 
  \#41& $2.50 \times 10^6$ & $-28.3$ & $5.57$  & $21$   \\ 
\end{tabular}
\caption{Parameters $t_0$, $\ln(\Delta v)$, $\tau$ and SNR for each injected glitch. For each glitch, we also indicate the corresponding number ($\#$) from the LPF catalogue.}
\label{tab:glitches}
\end{table}

\subsection{Data analysis\label{data_analysis}}

\noindent We analyse a complete 30-day dataset in a single run. Using the shapelet model with $n_0=1$, we fit for glitches, allowing up to 10 glitches in the data, also including the possibility of no glitches. The noise shape is considered fixed, while the two amplitude parameters of the OMS and TM noise, $N_{\rm OMS}$ and $N_{\rm TM}$, vary within the uniform prior range specified in Table~\ref{prior_on_glitch}. We also use uniform priors for the time of injection ($\tau_0$) and the decay time ($\beta_0$) of the shapelet, whereas for the amplitude ($A_0$)  of the shapelet, we consider a uniform prior in log amplitude. In this way, we avoid sampling parameters with vastly different numerical scales—such as amplitudes on the order of $10^{-12}$ and injection times around $10^4 \sim 10^6$. This choice tends to give more weight to low-SNR glitches. The goal here is to ensure that the sampler spends enough time in this region of parameter space, where most glitches occur. Figure~\ref{cap: threshold} illustrates this relationship by showing the impulse, duration, and SNRs of the glitches in the LPF catalogue. \\

\begin{figure}[ht!]
\centering
\includegraphics[width=\columnwidth]{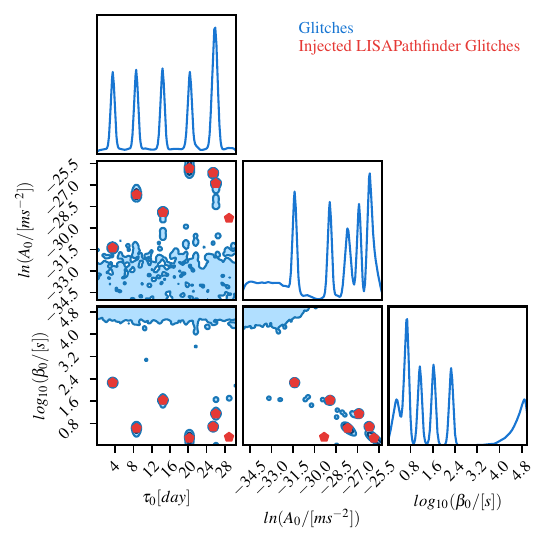}
\includegraphics[width=\columnwidth]{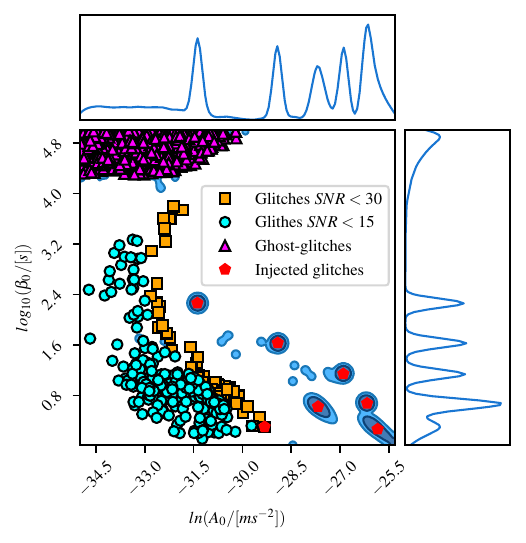}
 \caption{Upper plot: Posterior distributions of the shapelet parameters $\tau_0$, $\ln(A_0)$, and $\beta_0$ (plotted as $\log_{10}(\beta)$ for better visualisation) for the seven injected glitches, fitted using first-order shapelets.
Lower plot: Zoom-in on the 2D posteriors of $\ln(A_0)$ and $\beta_0$ (plotted as $\log_{10}(\beta_0)$), overlaid with SNR contours of the injected glitches.
\label{glitch_posterior}}
\end{figure}

\noindent The analysis takes about 2 days using 
70 walkers and 10 temperatures, running on a single A100 GPU. We expect that this timing can be improved in the future. Concerning convergence of the algorithm, it was assessed by monitoring the stability of the posterior distributions over successive blocks of 500 iterations. Specifically, we updated the covariance group move (see sec. \ref{glitch_in_model}) every 500 iterations and checked that the posteriors did not change significantly between consecutive blocks. In practice, convergence was often achieved earlier. We plan to investigate alternative stopping criteria in future work.
\begin{table}[h]
\begin{centering}
\begin{tabular}{|c|c|c|}
    \hline
   Parameter &  Lower Bound  & Upper Bound \\ 
    \hline
    $\tau_0$ [s]  & 0  & 2592000  \\ 
    \hline
    $ln(A_0/[m s^{-2}])$ & -35 & -20 \\ 
    \hline
      $\beta_0$ [s] & 1  & 1e5 \\  
      \hline
     $N_{\rm TM}$ [$m s^{-2}Hz^{-1/2}$]  & (7e-12) & (8e-12) \\
     \hline  
     $N_{\rm OMS}$ [$m Hz^{-1/2}$]  & (2e-15)  & (3e-15)  \\
     \hline
\end{tabular}
\caption{Table of priors used for shapelets and noise parameters.\label{prior_on_glitch}}
\end{centering}
\end{table}\\
Figure~\ref{glitch_posterior} reports the posterior for the parameters of the shapelets used to fit the injected glitches. 
The parameters of the injected glitches (Table \ref{tab:glitches}) --- converted to shapelet parameters according to $\tau_0 =t_0$, $A_0 = 
\frac{\Delta v}{2 \beta_0}$ and $\beta_0 = \tau$ --- are indicated within the Figure.

\begin{figure}[ht!]
 \centering
\includegraphics[width=\columnwidth]{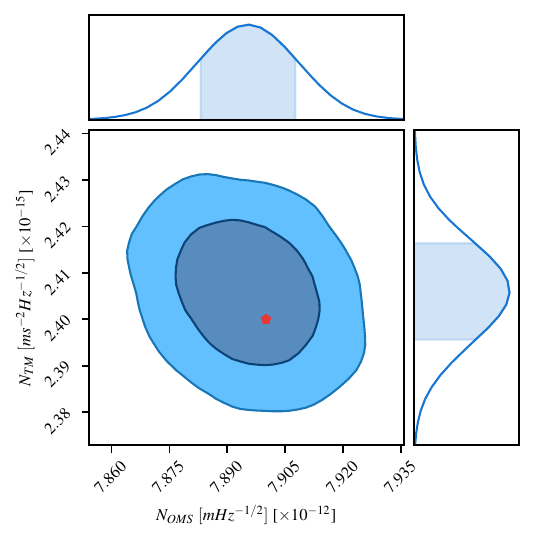}
 \caption{Posterior distribution of the amplitudes, $N_{\rm TM}$ and $N_{\rm OMS}$, for the test mass acceleration noise and optical metrology system noise\label{noise_posterior}}
\end{figure}
 \begin{figure}[h]
\includegraphics[width=\columnwidth]{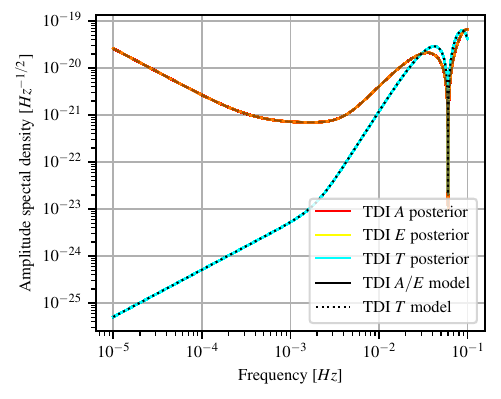}
 \caption{Amplitude spectral density (ASD) of the TDI $A,E$ and $T$ channels constructed from 1000 random draws from the posterior distribution of the amplitude noise parameters of test mass acceleration noise $N_{\rm TM}$, and optical metrology noise $N_{\rm OMS}$, after glitch-fitting is performed. The spread of the ASD across the posterior is very small, and the recovered ASDs are completely consistent with those used to generate the data}\label{noise_without_glitches}.
\end{figure}
\noindent It can be observed in Fig.~\ref{glitch_posterior} that the parameters of the glitch with the lowest SNR (21) cannot be recovered, and that the posteriors of the two glitches that are close in time overlap with each other such that in the 1D plot we observe a fifth bigger Gaussian for the time of injection $\tau_0$. The failure to recover the low-SNR glitch is likely due to the fact that the search did not detect it, possibly because proposed points rarely fall within the correct region of parameter space. This issue is probably linked to the choice of a uniform prior on $\beta_0$ (along with its lower amplitude), which turns out to be too broad relative to the $\tau_0$  values for some of the injected glitches —despite this choice being consistent with the general LPF decay time distribution shown in Fig.~\ref{cap: threshold}. To verify this hypothesis, we repeated the analysis  using a logarithmic prior on $\beta_0$ and we observed that the low-SNR glitch is better constrained\footnote{Additionally, to exclude the possibility that the low-SNR glitch is consistent with noise, we have performed another test in which we fixed the number of glitches to seven, initialized the sampler at the true parameters and allowed the sampler to adjust slightly. Once the parameters stabilized, we restarted the run with RJ turned on, allowing the sampler to remove glitches. The sampler did not remove any of the glitches once they were found, indicating that—at least under the specific settings used in this example—this low SNR glitch is not consistent with noise.}. However, it is important to underline that the recovery of the two noise parameters is nonetheless unbiased, as shown in Fig.~\ref{noise_posterior}. Therefore, the recovered PSD is well-constrained (Fig.~\ref{noise_without_glitches}) and closely aligns with the PSD model used to simulate the data, 
as illustrated in Fig.~\ref{noise_without_glitches}. In the lower plot of Fig.~\ref{glitch_posterior}, we also observe a larger number of fictitious low-amplitude glitches (``ghost glitches'') arising from random fluctuations in the stationary component of the noise. These glitches have SNRs $\lesssim 24$. We also plot the glitches from LPF with $\rm SNR < 15$ and $15 < \rm SNR < 30$. Notably, the low-SNR glitch with $\rm SNR = 21$, which was not fitted, lies within the region of the latter. This suggests that glitches with parameters in the same region as those are consistent with noise, given our choice of prior.

\begin{figure}[ht]
\includegraphics[width=\columnwidth]{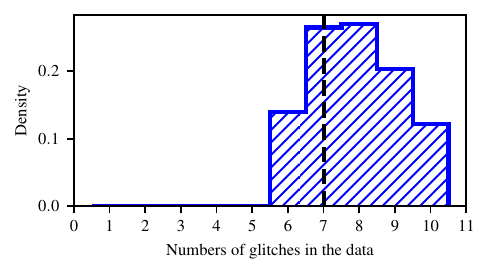}
 \caption{
 Distribution of the number of glitches at zero temperature estimated by the dynamical parameter estimation procedure. 
 The true value is marked with a dashed black line. \label{cap:num_of_glitch}}
\end{figure}

\noindent Finally, Fig. \ref{cap:num_of_glitch} shows the number of estimated glitches in the data at zero temperature. Even though seven glitches were injected, the sampler prefers to have eight glitches, which means adding two ``ghost glitches", since one of the injected ones is not recovered. The key difference between the ``ghost glitches'' and the missing one is that the amplitudes of the former are so low that the associated sampled value of $\beta_0$ and $\tau_0$ have no impact on the likelihood, and so, if such signals have significant prior support, they will readily be added and removed by the RJ step of the sampler. The amplitude of the missing glitch is small enough to evade recovery, but no so low that the associated $\tau_0$ and $\beta_0$ have no impact on the likelihood. Therefore, only values of $\tau_0$ and $\beta_0$ in the right portion of the parameter space give high enough likelihood, and glitches with amplitude similar to that of the missing glitch are less likely to be accepted. The addition of ``ghost glitches'' is the result of allowing for low-amplitude glitches in the prior, which can be used to fit random excursions in the instrumental noise. 
As these fictitious glitches do not absorb much of the power in the data, this does not impact noise estimation, but nonetheless, there are techniques to mitigate this effect. One approach, used for example in \cite{Littenberg_2020} in the context of including white dwarf binaries in the LISA Global Fit, is to use a stricter prior on the SNR of the white dwarf binaries that are proposed to be added.
\\

\noindent To estimate the lowest SNR glitch that can be distinguished from noise, we considered different scenarios in which the $\#41$ glitch, the low SNR one that was not recovered in our analysis, was replaced by other glitches of increasing SNRs starting from SNR 20 to 65. We concluded that, for the noise model used in this study, glitches with an SNR of a few tens cannot be distinguished from noise. In the next section we show that such glitches do not lead to significant biases in MBHB parameter estimation. Biases start to appear from SNRs of $\sim60$. This observation could also be used to define a prior for glitches based on their SNR to help mitigate the confusion noise from the low-SNR sources discussed in the preceding paragraph.

\section{Effect of glitches on parameter estimation of a Massive Black Holes Binary \label{sec:estimation_part}}
\noindent We consider now three different cases where we inject three glitches with SNRs of 21, 72 and 1544, respectively. The first glitch corresponds to $\#41$, the second to $\#3$ and the third to $\#248$ of the LPF catalogue. The parameters of these glitches are reported in the Table \ref{tab:glitches}.
We inject one glitch at a time, substituting it with another in each subsequent example. We simulate for all three cases a fiducial MBHB signal at redshift 4,  with SNR 1152, in noisy data. The choice of the MBHB is driven by Fig. 3.5 of \cite{colpi2024lisadefinitionstudyreport}, which shows the typical SNRs and redshift of MBHB sources expected in LISA, as well as Fig. 3 in \cite{amaroseoane2017laserinterferometerspaceantenna}.\\

\noindent The MBHB priors used for this analysis are given in Table ~\ref{mbmb_table}. Uniform distributions are used on parameters $\{\rm log_{10}(M_t), q, a_1, a_2, d_L, \varphi_{ref}, \lambda_d, \psi, t_{ref}\}$. The prior on total mass is log-uniform, and we express the masses as dimensionless multiples of the solar mass $\rm M_{\odot}$. For the inclination, we use a uniform prior in $\cos \iota$ between -1 and 1, while the prior on the ecliptic latitude is uniform in $\sin \beta_d$.\\
\begin{table}
\begin{centering}
\begin{tabular}{|c|c|c|c|}
    \hline
   Parameter & True & Lower Bound  & Upper Bound \\ 
    \hline
    $\rm log_{10}(M_t)$ & 6.29 & 5 & 8 \\ 
    \hline
    $q$ & 0.4628 &0.01 & 0.999 \\ 
    \hline
      $a_1$ &0.7474 & -1 & 1 \\  
      \hline
     $a_2$ &0.8388 & -1 & 1 \\  
     \hline
      $d_L [Gpc]$ & 37 & 0.01 & $1000$ \\  
 \hline
      $\varphi_{ref}$ $[rad]$ & 1.194 & 0 & 2$ \pi $\\  
\hline
      $\cos \iota$ & 0.5 & -1 & 1 \\  
    \hline
      $\lambda_d$ $[rad]$ & 1.288 & 0 & 2 $\pi$ \\  
\hline
      $\sin \beta_d$ &-0.298& -1 & 1 \\  
 \hline
   $\psi$ $[rad]$& $\pi/6$ & 0 & $\pi$ \\  
  \hline
   $t_{ref} [s]$ & $2.628\times10^6$ & $t_{ref} $ - 100 & $t_{ref}$ + 100 \\  
   \hline
\end{tabular}
\end{centering} \caption{True values of the MBHB parameters together with the prior range used for parameter estimations. We assume a flat prior on the specified range. Values of the masses are given in the detector frame\label{mbmb_table}.}
\end{table}

\noindent We choose the injection time for each of the three glitches to maximize the impact of the glitch on the MBHB parameter estimation, following the procedure described in \cite{Spadaro_2023}. The idea is to maximize the overlap between the glitch and the MBHB signal.
The noise-weighted inner product can be defined as:
\begin{equation}
  (h_1| h_2) = 4 \Re \int_0^\infty h^{\dagger}_1(f) \, \mathbf{C}^{-1}(f) \,h_2(f) \, df,
\end{equation}
Here, $h_1(f)$ and $h_2(f)$, respectively, stand for the vectors of MBHB signals and glitches as they appear in the $N_{\rm TDI}$ TDI combinations considered in our analysis. $\mathbf{C}(f)$ is the cross-spectral density matrix of the noise in those channels and $\dagger$ denotes the Hermitian transpose. Using this inner product, the normalized overlap is given by:
\begin{equation}
  \mathcal{O}(h_1, h_2) = \frac{(h_1 | h_2)}{\sqrt{(h_1 | h_1)(h_2 | h_2)}}.
\end{equation}
When considering multiple independent channels, for example, TDI $A$ and $E$, the inner product reduces to a sum over the channels:
\begin{equation}
  (h_1 | h_2) = 4 \Re \int_0^\infty \sum_{j=1}^{N_{\rm TDI}} \frac{h_1^{j*}(f) \, h_2^{j}(f)}{S_n(f)} \, df,\label{eq:maximum_overlap}
\end{equation}
where we consider the $A$ and $E$ channels to not only be uncorrelated but also to have equal noise PSD, $S_n(f)$.  \\

\begin{figure}[ht]
\centering
\includegraphics[width=\columnwidth]{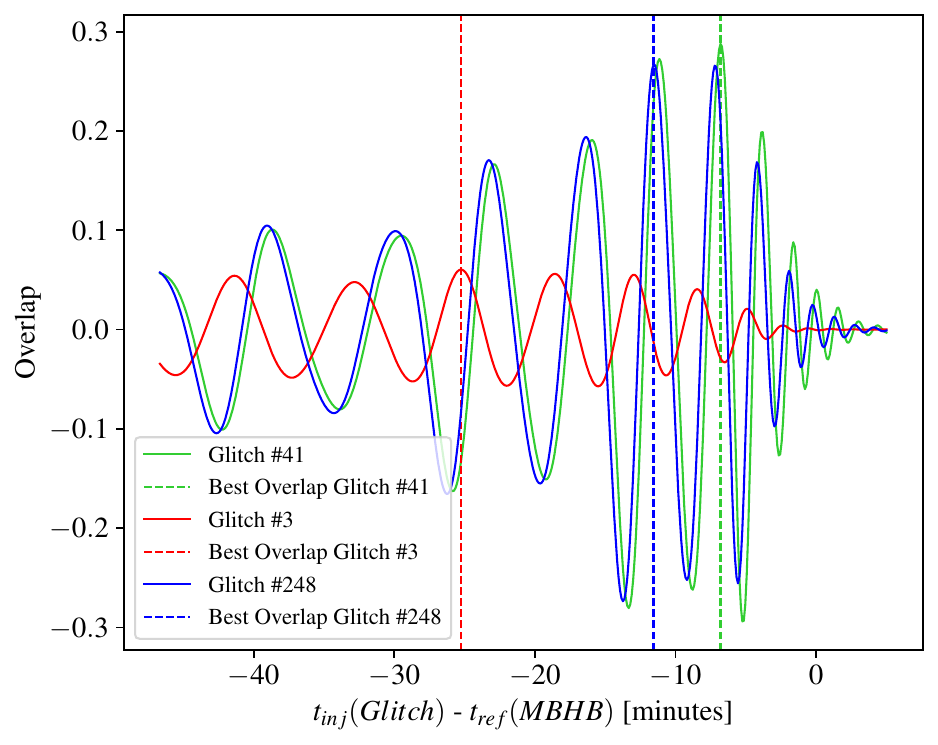}
\caption{Overlap between the MBHB and glitch signals as a function of the glitch injection time relative to the merger, where the decay time and amplitude of the glitches are kept fixed. The red, blue, and green solid curves are for the three glitches with SNRs of 21, 72 and 1544. The GW signal is fixed to that of our fiducial MBHB. Dashed vertical lines with matching colours denote the times that maximize the overlap.\label{Fig:max_matching}}
\end{figure}

\begin{figure}[ht!]
\centering
\includegraphics[width=\columnwidth]
{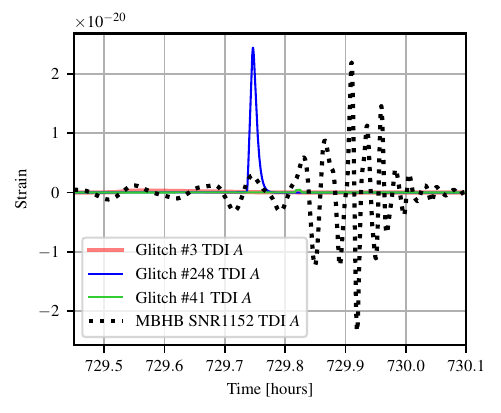}
\caption{Fiducial MBHB waveform along with the  glitches injected at their maximum overlap time in TDI channel $A$.\label{Fig:glitch_mbh}}
\end{figure}

\noindent Defining $\Delta t$ the difference between the MBHB merger time and the glitch injection time, we found that the best match for the glitch with SNR 21 is at $|\Delta t|=6.80361$ minutes before the merger, for the glitch with SNR 72 it is at $|\Delta t| =25.2338$ minutes before the merger, and for the glitch with SNR 1544 at $|\Delta t|=11.5665$ minutes before the merger. Figure \ref{Fig:max_matching} shows the variation of the Overlap versus time for the three glitches, where we have fixed the amplitudes and decay time of the glitches while varying the injection time. The dotted vertical lines indicate the injection time, with respect to the merger, at which the overlap is maximised for each of the three glitches. We show in Fig. \ref{Fig:glitch_mbh} the three glitches injected at the time that maximizes the Overlap between the glitch and the fiducial MBHB in TDI channel $A$.\\ Note that the frequency of the MBHB evolves over time, and the maximum impact of a glitch is expected when its frequency overlaps with the instantaneous frequency of the MBHB signal. Acceleration glitches typically occur at frequencies on the order of $10^{-3}$ Hz, which aligns more closely with the early inspiral phase. However, if the MBHB merger occurred at lower frequencies—as in the case studied in \cite{Spadaro_2023}—the glitch would instead have the strongest overlap during the merger phase.

\begin{figure*}
\centering
\includegraphics[width=\textwidth]{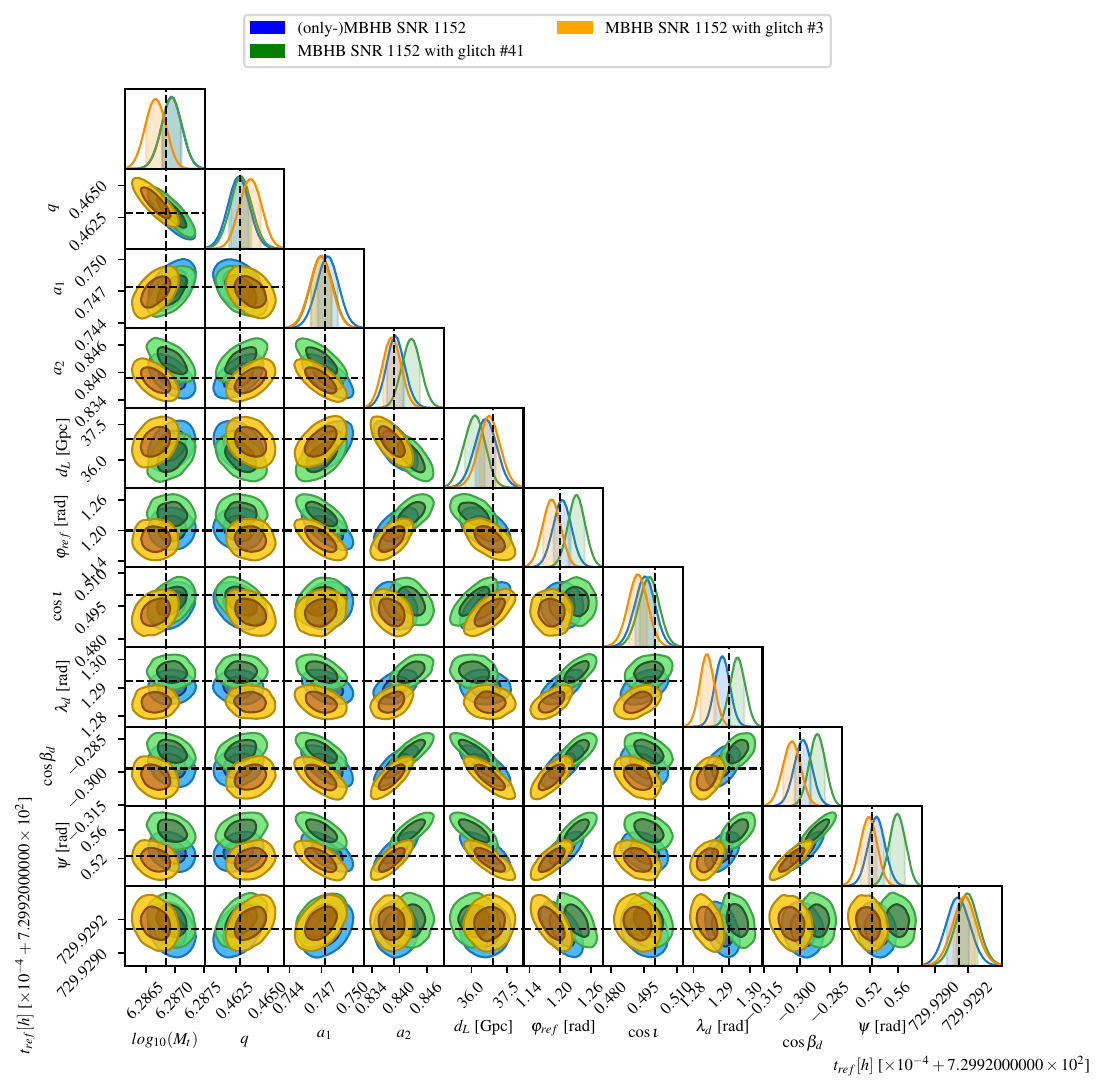}
\caption{Posteriors of an MBHB with SNR 1152 compared to two cases: a glitch of SNR 21 is injected at the maximum match but not fitted, a glitch of SNR 72 is injected at the maximum match but not fitted. \label{fig:corner_plot_mbhb_glitches}}
\end{figure*}

\begin{figure}
\centering
\includegraphics[width=\columnwidth]{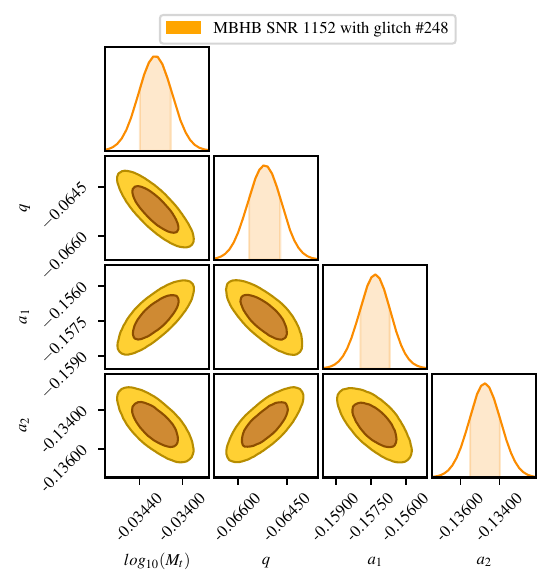}
\caption{Posteriors of the intrinsic parameters of an MBHB with SNR 1152 in the case where a glitch of SNR 1544 is injected -and not fitted - at the time maximising its match with the injected signal. We have shifted the parameters relative to the injected values. The fact that the posterior do not peak at 0 illustrates the bias caused by the glitch. For the sake of clarity, we do show the corresponding posteriors for the only-MBHB case, as the statistical error is much smaller than the systematic bias observed in the figure.\label{fig:corner_plot_mbhb_glitch}}
\end{figure}

\noindent The posteriors shown in Fig.~\ref{fig:corner_plot_mbhb_glitches} and Fig.~\ref{fig:corner_plot_mbhb_glitch} illustrate 
, the effect of the presence of these different glitches if we attempt to recover the parameters of this MBHB without simultaneously fitting for both the artefact and the GW source. For comparison, the posteriors of the MBHB parameters obtained from data containing no glitches are reported in both plots, and the true values of the parameters are indicated by a dashed black line. \\

\noindent In the case of the SNR 21 glitch, shown in Fig~\ref{fig:corner_plot_mbhb_glitches}, the glitch has less impact in the intrinsic parameters with respect to the impact we see in the extrinsic ones, as the shift in the posterior is less than the width of the posterior. The bias starts to be significant for the glitch with SNR 72 for several of the parameters, e.g. the total mass, the reference phase at merger and the ecliptic longitude and latitude. For the SNR 1544 glitch shown in Fig.~\ref{fig:corner_plot_mbhb_glitch}, the two posteriors are completely disjoint, with all parameters significantly biased. To help the reader visualise the effect of not fitting such a glitch, we report only the posteriors for the intrinsic parameters. It is important to specify that the noise model has been fixed for this analysis, and we expect that variations in the noise model would allow to partially absorb the impact of the glitch 
making the MBHB posterior more consistent with the injected values. \\

\noindent Let us recall that the injection times of the glitches were chosen specifically to maximise their impact on the parameter estimation of the MBHB. Their effect is expected to be smaller if they occur during the early inspiral. This point is further discussed in Appendix~\ref{sec:mbhb_glitch}, where we present posteriors illustrating how the presence of the SNR~1544 glitch affects parameter estimation for the same MBHB, depending on its injection time.
However, since the coalescence is the loudest part of MBHB signals and largely drives parameter estimation, failing to capture glitches occurring near to it would severely jeopardise the science objectives of LISA. For instance, biases in sky localisation and distance could prevent the detection of an electromagnetic counterpart and lead to misidentification of the host galaxy, thereby compromising cosmological measurements based on bright or dark siren methods~\cite{Gair:2022zsa,Caprini:2016qxs,Tamanini:2016zlh,Mangiagli:2023ize}.
This issue also extends to applications in fundamental physics, which rely on detecting extremely small deviations from general relativity. In this context, even small biases can lead to false detections and must be mitigated as much as possible \cite{Kwok:2021zny,Gupta_2025}.
Our analysis shows that, to avoid such pitfalls, glitches with SNRs of several tens or higher will need to be included in the model fitted to the data. This threshold also corresponds to the point at which glitches first become identifiable in the data, as discussed in the previous paragraph.


\begin{figure*}
\centering
\includegraphics[width=\textwidth]{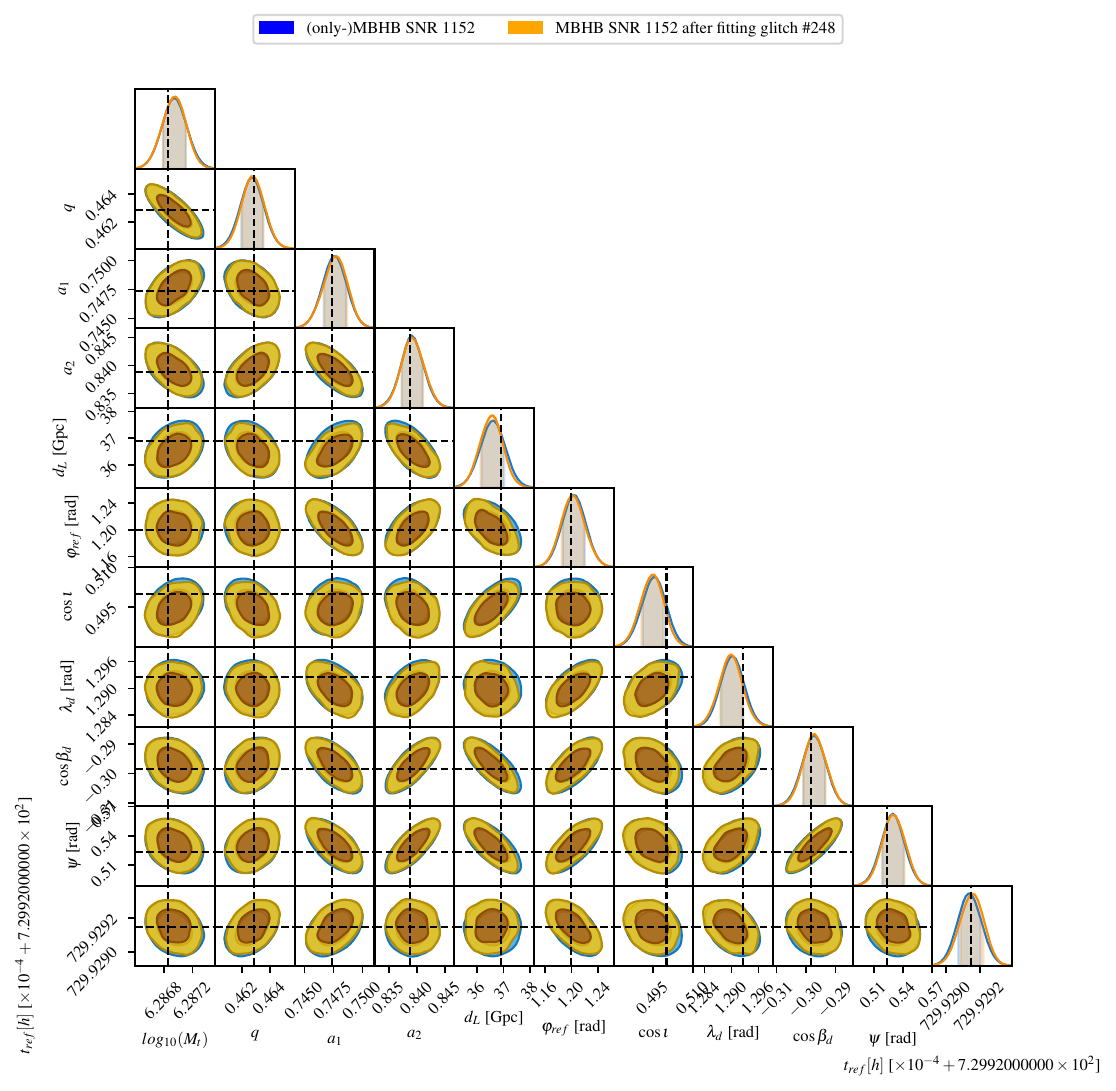}
\caption{Posteriors for the parameters of an MBHB with SNR 1152 obtained from data containing only the MBHB, compared to those obtained from data containing a glitch of SNR 1544, when the latter is also fitted.
\label{fig:corner_plot_mbhb_glitch_fitting}}
\end{figure*}

\begin{figure}
\centering
\includegraphics[width=\columnwidth]{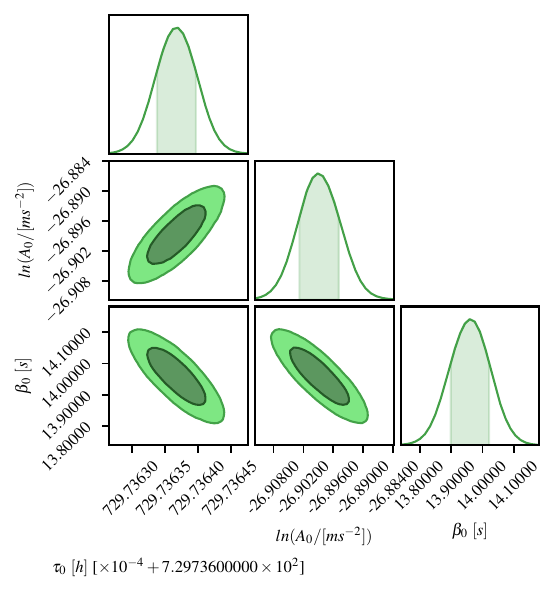}
\caption{Posteriors on the parameters of the first order shapelet used to fit the injected glitch ($\#248$) with SNR 1544. \label{fig:corner_plot_glitch_fitting}}
\end{figure}

\noindent In Fig.~\ref{fig:corner_plot_mbhb_glitch_fitting}, we now show the posterior on the parameters of the MBHB signal obtained when simultaneously fitting for the glitch, and compare this to the posterior obtained in the absence of the glitch. We observe that by fitting the glitch, the MBHB posteriors are no longer biased, and the posteriors for the (only-)MBHB case overlap with those where the glitch has been accounted for, indicating that the best strategy is to fit for both. For completeness, we also show posteriors for the fitted glitch parameters in Fig.~\ref{fig:corner_plot_glitch_fitting}.   \\

\noindent We investigated the effect of injecting glitches ($\#41$) and ($\#3$) into MBHB with a SNR of 320, in order to reproduce the example studied in \cite{Spadaro_2023}. The posteriors plots are reported in the Appendix \ref{sec:mbhb_glitch_second_case}. For the short-duration glitch with SNR 21 ($\#41$), we find results consistent with those reported there: the inferred parameters remain largely unbiased. In contrast, glitch ($\#3$), with an SNR of 72, though also short in duration, introduces a more significant bias.

\section{``(Light-)\textit{Spritz}'' data set}\label{sec:spritz_data}
\noindent In this section, we describe the ``Spritz'' data set and the strategy we have used to analyse it. We describe the mock data in section \ref{sec:spritz-intro} and the noise model used for this analysis in section \ref{sec:spritz-noise}, since it differs from the noise that was used to generate and analyse the data in the previous section \ref{sec:estimation_part}. Finally, in section~\ref{sec:strategy}, we present the methodology we use for doing the initial search and parameter estimation of glitches, noise and MBHB and for the subsequent analysis of the full data set to refine the posterior estimates of the noise, GW signal and artefacts.  

\subsection{Introduction}\label{sec:spritz-intro}
\begin{figure}
\centering
\includegraphics[width=\columnwidth]{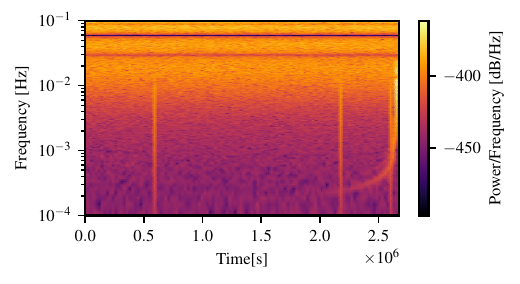}
\caption{Time-frequency representation of the ``\textit{Light-}Spritz'' data set\label{fig:time_frequency_spritz}. }
\end{figure}

\begin{figure}
\centering
\includegraphics[width=\columnwidth]{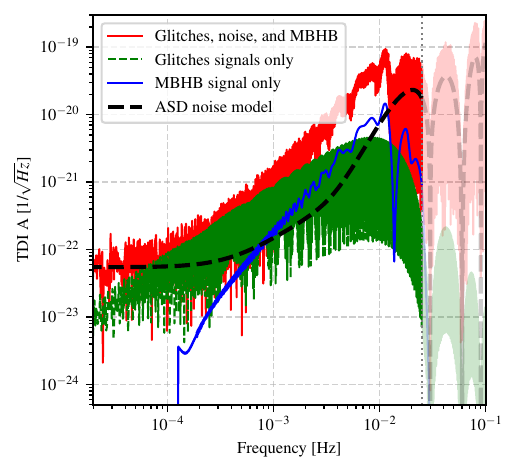}
\caption{Fourier domain representation of the ``\textit{Light-}Spritz'' data set\label{fig:FFT_spritz}.}
\end{figure}

\noindent Full details of the ``Spritz'' data-sets can be found in~\cite{LDCS}. The ``Spritz'' data release comprised three main data sets, of which we focus on the first one, which contains a single loud MBHB and three glitches as visible in Fig.~\ref{fig:time_frequency_spritz}, which shows the dataset in the time-frequency domain and  Fig.~\ref{fig:FFT_spritz} which shows it in the frequency domain. The glitches are so-called double decay exponential glitches~\cite{Sala:2023hpr}. All the data sets are realised with TDI $X,Y$ and $Z$ second generation assuming Keplerian orbits. 
The dataset also features more realistic instrumental noise (see \cref{sec:spritz-noise}), a background of Galactic binaries and data gaps. Auxiliary data included in the dataset allows individual components to be perfectly subtracted, allowing analysts to focus on selected aspects of the dataset. Since we focus on the detection of glitches and their impact on MBHB parameter estimation, we removed the gaps and Galactic binaries, and for this reason, we renamed the modified data set as ``\textit{Light}-Spritz''. Furthermore, the version of BBHx we are using for this paper (version 0.0.0) does not support the response model used to generate the MBHB in ``Spritz''. The ``Spritz'' data was generated using the PhenomD waveform model \cite{Husa_2016, Khan_2016}, with the LISA response evaluated for Keplerian orbits (time varying arm-lengths), but our version of BBHx does not include time varying arm-lengths response. Therefore, to rule out any systematic errors from this waveform mismatch~\cite{dhani2024systematicbiasesestimatingproperties}, we replace the original ``Spritz'' injected signal with one with the same parameters (see Table \ref{mbmb_table} in sec.~\ref{sec:estimation_part}), but generated using the PhenomHM waveform (instead of PhenomD)
with an equal arm response. We plan to
include a more realistic treatment of the LISA constellation, to compute the response, in a follow-up study that will analyse the full data set.

\noindent 
\noindent Regarding glitch modelling, Ref.~\cite{Muratore:2021rwq} showed that the mismatch between equal- and unequal/time-varying-arm models does not significantly impact glitch parameter reconstruction. For the noise, however, using equal, unequal, or time-varying arm-length models affects the high-frequency range, where the zeros of the TDI appear. To avoid related issues, we restrict our analysis to frequencies up to \SI{2e-2}{\hertz} \cite{Hartwig_2022}, which does not alter the outcome of our analysis. Additionally, the null channel $T$ cannot be well approximated using the equal-arm-length model~\cite{PhysRevD.105.023009,PhysRevD.82.022002}. For these reasons, we perform the analysis using only the two sensitive channels, $A$ and $E$, employing second-generation TDI under the assumption of equal arm lengths, for which sensitivity to discrepancies in arm-length modelling is negligible. For consistency, the MBHB signal is also computed using second-generation TDI variables (see Eq.~\eqref{eq:tdi2-definition}), assuming constant and equal arm lengths.\\

\noindent As discussed in Sec. \ref{sec:likelihood-formulation}, we analyse the data in the frequency domain. However, converting the time-domain data provided in ``\textit{Light-}Spritz'' to the frequency domain by using a fast-Fourier transform (FFT) without any pre-processing causes significant artefacts in the form of spectral leakage \cite{Hartwig:2021dlc}, linked to significant high-frequency content that leaks into the low frequency part of the ``\textit{Light-}Spritz'' data set. We mitigate this effect by applying a low-pass filter to the data before taking the FFT. Another approach would be to apply a window function to the data before taking the FFT. This has the downside of introducing correlations between adjacent frequency bins, potentially biasing the parameter estimation using our diagonal likelihood model. Instead, by filtering the data, we obtain a diagonal likelihood (see Appendix \ref{app:filter discussion}).
Concretely, we use a first-order Butterworth filter with a cut-off frequency of \SI{1e-3}{\hertz} using the \texttt{scipy.signal.butter} function and then use the \texttt{scipy.signal.filtfilt} function to apply the filter twice in opposite time directions, such that it has zero impact on the phase of the underlying data (i.e., it does not cause any time-shift of the data). Moreover, in order not to bias our analysis, we apply a frequency domain model of the filter response to the noise, glitches and GW signal templates used in the likelihood as well. Due to the double application of the filter, the Fourier transforms appearing in the likelihood are scaled by $|H(f)|^2$, whereas PSDs are scaled by $|H(f)|^4$, where $H(f)$ is the complex filter transfer function: 

{\small
\begin{align}
& \log \mathcal{L} = \nonumber \\
 & -\frac{1}{2} \sum_{f} \Bigg[ 
    4 \Delta f \, \frac{
        \left| 
            \tilde{d}(f) 
            - \tilde{h}(f, \theta) |H(f)|^2 
            - \tilde{g}(f, \phi) |H(f)|^2 
        \right|^2 
    }{
        S(f, \gamma) |H(f)|^4
    } \nonumber \\
& \qquad\qquad 
+ \log \left\{ S(f, \gamma) |H(f)|^4 \right\} 
\Bigg].
\end{align}
}

\subsection{Noise model}\label{sec:spritz-noise}
\noindent To study -- and model -- just the noise in the ``\textit{Light-}Spritz'' dataset, we further removed the glitch and MBHB components described in \cref{sec:spritz-intro}. We then estimated the spectra using the \texttt{scipy.signal.welch} method with the following parameters: \texttt{window=('kaiser', 10)}, \texttt{nperseg=10713},  \texttt{noverlap=5356}, \texttt{detrend='linear'}. The LDC manual for ``Spritz''~\cite{LDCSA} suggests that laser, backlink, acceleration and readout noises were enabled in the simulation. However, we find that the corresponding noise model provided as part of the LDC software suite (\cite{lisa_data_challenge_working_group_2022_7332221}, v.1.2.4) does not perfectly describe the ``Spritz'' simulated data. \\

\noindent To accurately fit the noise of the data set, we cross-checked against the default values for the noise sources included in the instrument simulation at the time~\cite{bayle_2022_6461078} and consequently included the following noise components, expressed as PSDs:
\begin{itemize}
    \item Inter-spacecraft interferometer OMS noise,\\ $S_{\mathrm{OMS}}^{\mathrm{ISI}} = \left( N_{\rm OMS}^{\rm ISI} \right)^2 $ (white),\\
    with $N_{\rm OMS}^{\rm ISI} = \SI{6.35}{\pico\meter\per\sqrt\hertz}$
    \item Reference interferometer OMS noise, \\ $S_{\mathrm{OMS}}^{\mathrm{RFI}} = \left( N_{\rm OMS}^{\rm RSI}\right)^2 $ (white),\\
    with $N_{\rm OMS}^{\rm RSI} = \SI{3.32}{\pico\meter\per\sqrt\hertz}$
    \item Test-mass interferometer OMS noise, \\ $S_{\mathrm{OMS}}^{\mathrm{TMI}}=\left( N_{\rm OMS}^{\rm TMI} \right)^2  $ (white),\\
    with $N_{\rm OMS}^{\rm TMI} = \SI{1.42}{\pico\meter\per\sqrt\hertz}$
    \item TM acceleration noise, \\ $S_{\mathrm{TM}}=\left( N_{\rm TM} \right)^2 \cdot \left( 1 + \left( \frac{0.4 \times 10^{-3}}{f} \right)^2 \right)$,\\
      with $N_{\rm TM} =\SI{2.4}{\femto\meter\per\second\squared\per\sqrt\hertz}$
    \item RFI and TMI backlink noise, \\ $S_{\mathrm{BL}}^{\mathrm{RFI/TMI}} = \left( N_{\rm BL}^{\mathrm{RFI/TMI}}  \right)^2 \cdot \left( 1 + \left( \frac{2 \times 10^{-3}}{f} \right)^4 \right)$,\\
    with $N_{\rm BL}^{\mathrm{RFI/TMI}} =\SI{3}{\pico\meter\per\sqrt\hertz}$.
\end{itemize}

\noindent For the analysis, we define the following combinations of the noise parameters to estimate the effective noise levels:
\begin{equation}
\gamma = \begin{cases}
N_1 \equiv \left[(N_{\mathrm{OMS}}^{\mathrm{ISI}})^2 + (N_{\mathrm{OMS}}^{\mathrm{RFI}})^2 + (N_{\mathrm{BL}}^{\mathrm{RFI/TMI}})^2\right]^{1/2} , \\
N_2 \equiv \left[(N_{\mathrm{OMS}}^{\mathrm{TMI}})^2 + (N_{\mathrm{BL}}^{\mathrm{RFI/TMI}})^2\right]^{1/2} , \\
N_{\mathrm{TM}} ,
\end{cases}
\end{equation}
with $N_1 =  3.32 \times 10^{-12}$, $N_2= 7.78 \times 10^{-12}$, and $N_{\mathrm{TM}} = 2.4 \times 10^{-15}$. As in \cref{noise_model}, we first express the noises as equivalent displacement noise in \si{\meter\squared\per\hertz} or acceleration noise in \si{\meter\squared\per\second^4 \per\hertz}. We then convert them to units of fractional frequency fluctuations as used in the LDC by applying the factors  $C_{\mathrm{disp}}^{\mathrm{ffd}} = \left({2 \pi f}/{c}\right)^2$ and $C_{\mathrm{acc}}^{\mathrm{ffd}} = \left({c 2 \pi f}\right)^{-2}$, where $c$ is again the speed of light. \\

\begin{figure}
\includegraphics[width=\columnwidth]{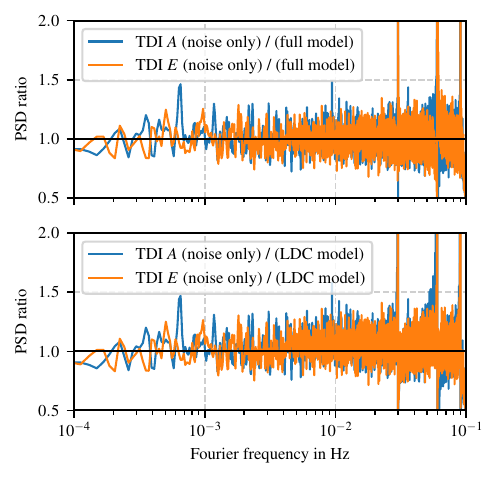}
\caption{Comparison between the power spectral density of the ``Spritz'' noise data set with the noise model reported in the  LDC manual.\label{fig:noise_model_spritz}}
\end{figure}

\noindent Fig.~\ref{fig:noise_model_spritz} shows the ratio between the estimated spectra (the PSD computed from the ``Spritz'' only-noise data), the model included with the LDC tools, and the model we used. We see artefacts around the zeros of the TDI transfer function for both plots, owing to the use of an equal arm approximation in the model we are using to fit the noise. For our model, we estimated the armlength to be used in the equal arm model as an average over all six time-varying armlengths, evaluating to $L / c \approx 8.323 s$. Whereas we used the default value set in the LDC tools for the LDC model. This causes slightly increased artefacts for the LDC model around the zeros. We also observe that the LDC model slightly underestimates the noise in the upper half of the frequency band. 
\\

\noindent To conclude, some of these noises have different TDI transfer functions, which are derived in \cite{Quang_Nam_2023} for TDI $X, Y, Z$ as well as $A, E, T$. For our work, we only need the transfer functions for $A$ and $E$, which turn out to be identical under our assumptions. We summarise them here for convenience. They include a common TDI factor for second-generation TDI:
\begin{equation}
C_{xx} = 16 \sin ^2(2 \pi  f L) \sin ^2(4 \pi  f L).
\end{equation}
Using this, we have
\begin{align}
    |H_{\text{ISI/RFI (OMS)}}^{AA}|^2 &= 2 C_{xx} \left( 2 + \cos(2 \pi f L) \right) \\
    |H_{\text{TMI (OMS)}}^{AA}& |^2  = \nonumber \\ & C_{xx} \left( 3 + 2 \cos(2 \pi f L) + \cos(4 \pi f L) \right) \\
    |H_{\text{TM}}^{AA}|^2 &= 4 \cdot |H_{\text{TMI (OMS)}}^{AA}|^2,
\end{align}
such that the total noise can be described by
\begin{equation}
\begin{split}
    S_{\rm AA}^{\mathrm{total}} & = S_{\rm EE}^{\mathrm{total}} = C_{\mathrm{disp}}^{\mathrm{ffd}} \Big[ 
    |H_{\mathrm{TMI (OMS)}}^{\rm AA}|^2 \left( S_{\mathrm{OMS}}^{\mathrm{TMI}} + S_{\mathrm{BL}}^{\mathrm{TMI}}
    \right)  \\
    &+ 
    |H_{\mathrm{ISI/RFI (OMS)}}^{\rm AA}|^2 \left( S_{\mathrm{OMS}}^{\mathrm{ISI}} + S_{\mathrm{OMS}}^{\mathrm{RFI}} + S_{\mathrm{BL}}^{\mathrm{RFI}}
    \right)\Big] \\
    &+ C_{\mathrm{acc}}^{\mathrm{ffd}} |H_{\mathrm{TM}}^{\rm AA}|^2 S_{\mathrm{TM}}.
\end{split}
\end{equation}

\section{Analysis of the data set\label{sec:strategy}}

\begin{figure*}
\centering
\includegraphics[width=0.85\textwidth]{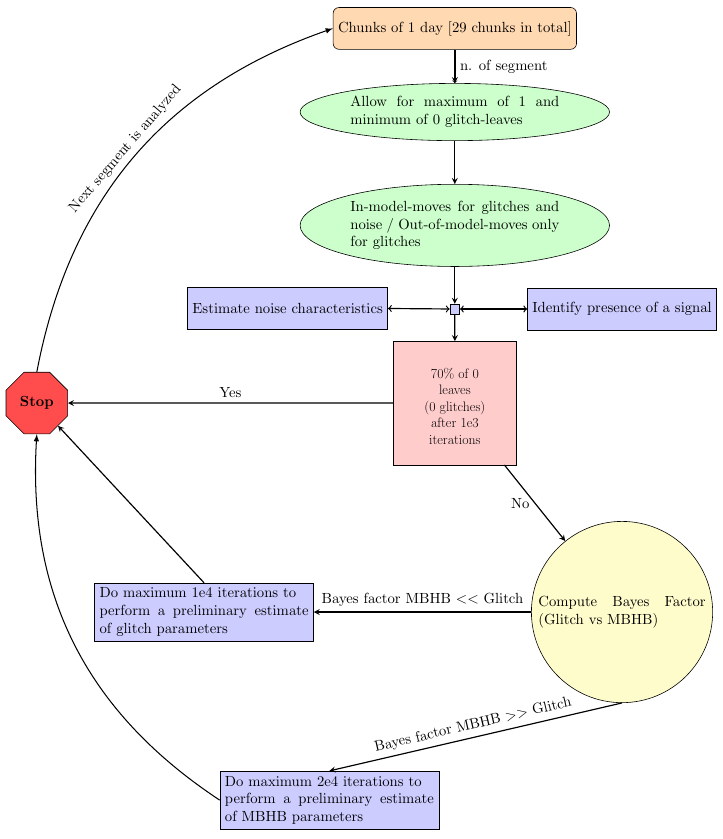}
\caption{Schematic view of the search algorithm.
\label{fig:search_phase}}
\end{figure*}

\begin{figure*}
\centering
\includegraphics[width=\textwidth]{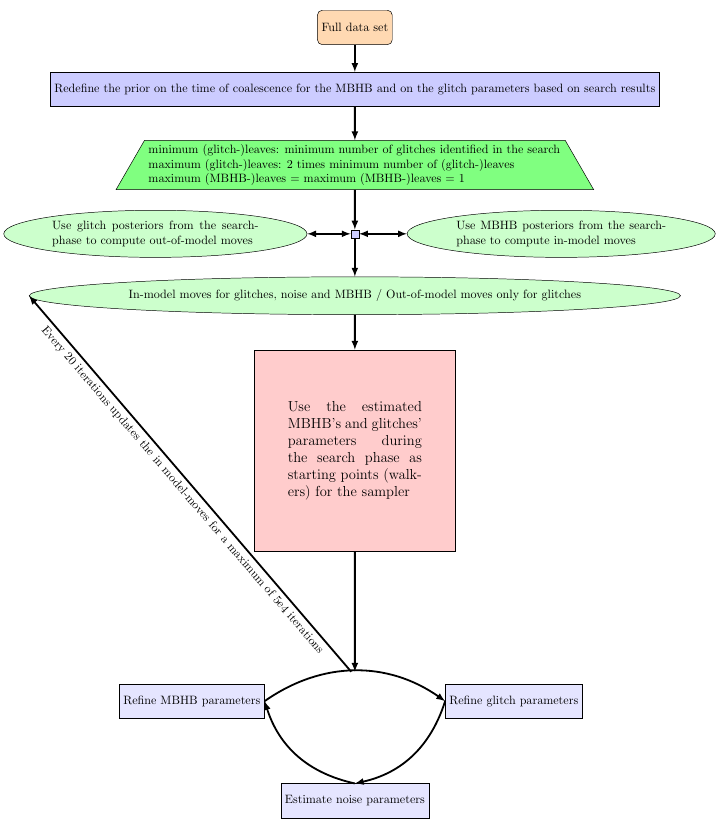}
\caption{Schematic view of the parameter estimation algorithm. \label{fig:PE_phase} }
\end{figure*}

\noindent The search-phase algorithm is visualized in Fig. \ref{fig:search_phase} and explained in the subsection \ref{search_phase}, while the parameter estimation one is visualized in Fig. \ref{fig:PE_phase} and explained in section \ref{pe_phase}.

\subsection{Search phase\label{search_phase}}
\noindent We divide the time series into segments of approximately one day, with a $20\%$ overlap between adjacent segments. We therefore segment the ``\textit{Light-}Spritz'' dataset into 29 stretches
of data and we regenerate a time grid spanning from $t_0 = \SI{0}{\second}$ to $t_{end}= \SI{2675890}{s}$ by subtracting the initial time of the ``Spritz'' dataset $t_{in}=\SI{8899200}{s}$.  We consider broad prior ranges for the glitch parameters, allowing the distribution of $A_0$ and $\beta_0$ to extend beyond what is observed in LPF (Fig. \ref{cap: threshold}), aiming to remain as agnostic as possible.\\
\noindent The priors and the corresponding true values on the glitch and noise parameters for the search phase are reported in Table~\ref{prior_on_glitch_search}. The true values of the amplitude and decay times for the glitch parameters are taken from the ``Spritz'' LDC manual \cite{LISA_LDC_Challenge2b}. In particular the amplitude $\Delta v$  is \SI{2.20}{\pico\meter\per\second}, while the two decay times are $\tau_1 = \SI{10}{\second}$ and $\tau_2 = \SI{11}{\second}$. The true value of the glitch injection times $\tau_{\rm inj}$ are provided in the ``Spritz'' dataset in days $\tau_{{\rm inj}_{gl_1,gl_2,gl_3}} = \{110, 128, 133\}$. We thus convert each of them to the time scale $\tau_{0_{gl_1,gl_2,gl_3}}$, used in our analysis, as
\[\tau_{0_{gl_1,gl_2,gl_3}} = \frac{\tau_{{\rm inj}_{gl_1,gl_2,gl_3}} \times 86400 - t_{\rm in} - n_{\rm skip} \times dt}{3600},
\]
where \(n_{\rm skip} = 1000\) is the number of samples skipped to account for the burn-in period of the filter. \\

\noindent Our search is split in two phases: $First$ $phase$ and $Second$ $phase$.
\begin{table}[ht!]
\begin{centering}
\begin{tabular}{|c|c|c|c|}
    \hline
    Parameter & True Value & Lower Bound & Upper Bound \\
    \hline
    $\tau_{0_{gl_1}}$ [s] & 166.61 & t-start & t-end \\
    $\tau_{0_{gl_2}}$ [s] & 598.61  & t-start & t-end \\
    $\tau_{0_{gl_3}}$ [s] &  718.61 & t-start & t-end  \\
     \hline
    $\ln(A_0/[m\,s^{-2}])$ & -27.58 & -35 & -20 \\
    \hline
    $\beta_0$ [s] & 10.5 & 1 & \num{1e5} \\
    \hline
    $N_1$ [$\si{\meter\per\sqrt\hertz}$] & \num{7.77e-12} & \num{7e-12} & \num{8e-12} \\
    \hline
    $N_2$ [$\si{\meter\per\sqrt\hertz}$] & \num{3.32e-12} & \num{2.5e-12} & \num{3.5e-12} \\
    \hline
    $N_{TM}$ [$\si{\meter\per\second\squared\per\sqrt\hertz}$] & \num{2.4e-15} & \num{2e-15} & \num{3e-15} \\
    \hline
\end{tabular}
\caption{Table of priors for the glitch and noise parameters along with the true values. The variables `t-start' and `t-end' change depending on the stretch considered. Note that the following relationships hold between a double decay exponential glitch and the shapelet model used to fit it: $A_0 = \Delta v \frac{1}{1+\tau_2/\tau_1}$ and $\beta_0 =\frac{ \tau_1 + \tau_2}{2}$. \label{prior_on_glitch_search}}
\end{centering}
\end{table}

\subsubsection{First phase\label{search_first_phase}}
\noindent First, we fit only for both noise and glitches in each segment separately. We use the in-model and out-of-model moves described in sec.~\ref{sec:moves}, to explore the parameter space. The number of glitches in each segment is estimated between 0 and 1 (i.e. the number of leaves in the glitch branch is 0 or 1). The search includes a stopping criterion such that if, after 1000 iterations, more than $70\%$ of the samples have no glitch leaves, we conclude that no glitches are present in the segment and proceed to analyse the next one. Otherwise, the segment is flagged as containing a possible glitch signal. \\

\noindent Table~\ref{parameter_streches} reports the results of the first phase of the search pipeline, showing in which stretches of data glitches have been found and their SNRs. Note that the SNR given in the table is estimated from the simultaneous fitting of glitches and noise. Specifically, to calculate the SNR, we compute the mean of the posterior distribution of SNR values for each data stretch during the search phase. This mean is obtained by drawing 3000 random samples from the posterior distribution of the glitch and noise parameters. However, the instrumental noise is not sufficiently constrained; therefore, these SNRs have to be considered as a rough estimate. Indeed, we observed that only a day of data is not enough to estimate and constrain all the noise parameters properly. This is particularly true in stretches of data where the late inspiral part of the MBHB is present (N26-N28). Therefore, we need to properly constrain the noise when processing the full month of data. This is further discussed in the Appendix \ref{search_results}. \\

\noindent To perform this initial search, we used 25 walkers and 10 temperatures, and the run for each segment takes about 20 minutes. \\
\begin{table}[ht!]
\centering
\begin{tabular}{|c|c|c|c|c|}
\hline
\textbf{N} & \textbf{Time [s]} & \textbf{\% of 0} & \textbf{\% of 1} & \textbf{Glitch}  \\
&  & \textbf{Leaves} & \textbf{Leaves} & \textbf{SNR}  \\
\hline
0  &  0.0 - 107995 & 86  & 13  & 0.60 \\ 
1  & 72000 - 197995  &  66  & 36  &  1.27\\
2  & 162000 - 287995  & 87  & 13  &  0.39\\ 
3  & 252000 - 377995  & 86  &  14  &  0.47\\ 
4  & 342000 - 467995  & 85  &  15  & 0.50 \\ 
5  & 432000 - 557995  & 83  & 16  &  0.54 \\
\rowcolor{red!30} 6  & 522000 - 647995  & 10  &  90  &  294 \\ 
7  & 612000 - 737995 &  58 &  42 &  1.28 \\
8  & 702000 - 827995 & 87  & 13  &  0.42 \\
9  & 792000 - 917995 &  87 & 13  &  0.43\\
10 & 882000 - 1007995 &  56 & 44  & 1.42 \\
11 & 972000 - 1097995  &  79 &  21  & 1.46   \\
12 & 1062000 - 1187995  & 85  & 15  & 0.51\\
13 & 1152000 - 1277995  & 84  &  15  &  0.52\\
14 & 1242000 - 1367995  & 86  &  14  &  0.45 \\
15 & 1332000 - 1457995  & 87  & 12  & 0.40  \\
16 & 1422000 - 1547995  & 87  & 13  & 0.45\\
17 & 1512000 - 1637995  & 80  & 20  &  0.65 \\
18 & 1602000 - 1727995 &  84 &  16  &  0.65 \\
19 & 1692000 - 1817995  & 86  &  14  &  0.48 \\
20 & 1782000 - 1907995  &  86 &  14  &  0.47  \\
21 & 1872000 - 1997995  & 85  & 15  &  3.34\\
22 & 1962000 - 2087995 &  85 &  15  &  0.44 \\
\rowcolor{red!30} 23 & 2052000 - 2177995  & 3  & 97  &  269.14  \\
\rowcolor{red!30} 24 & 2142000 - 2267995  & 6  &  93  & 265.68  \\
25 & 2232000 - 2357995 & 81  & 18  &  0.77 \\
\rowcolor{cyan!30}26 & 2322000 - 2447995  & 8  & 92  & 3.74 \\
\rowcolor{cyan!30}27 & 2412000 - 2537995  &  3 & 97  & 5.85  \\
\rowcolor{red!30} 28 & 2502000 - 2627995  & 1  & 98  &  188.99  \\
\rowcolor{yellow!50} 29 & 2592000 - 2675890 & 1  & 99  &  520.26 \\ 
\hline
\end{tabular}
\caption{
Table showing the number of leaves in each model and the glitch SNR for each stretch of data analysed. 
$N$ labels the segment considered. The ``\% of 0 or 1 leaf'' columns indicate the fraction of samples with 0 or 1 leaf, respectively. The orange lines indicate stretches of data where glitches were identified, the cyan lines mark regions where the inspiral phase of the MBHB becomes significant, and the yellow line highlights the stretch where the merger of the MBHB is present.
}
\label{parameter_streches}
\end{table}\\
We observe five stretches with high SNR signals. Of these, four have been identified as glitches (orange line) and one as an MBHB (yellow line). All these stretches have at least $90\%$ of samples with only one leaf. Additionally, two other stretches (cyan line) have more than $92\%$ of samples with one leaf, although their estimated signal SNRs are much lower. This occurs because the inspiral phase of the MBHB waveform starts to contribute a non-negligible SNR, triggering the algorithm with evidence of a low-SNR signal. In the remaining stretches, the SNR values result from random noise fluctuation.  \\

\noindent In the first phase of the search, the glitch model serves as a trigger to detect the presence of a signal in each stretch, whether it corresponds to a glitch or an MBHB signal. This approach is effective because the MBHB in ``\textit{Light-}Spritz'' has a high SNR, leading to behaviour similar to matched filtering, where the overlap between our template bank and the GW signal increases with the SNR (the characteristic frequency of the glitch and the instantaneous GW frequency of the MBHB). However, in cases with lower SNR MBHBs, the search may not be sensitive enough to detect such signals. To resolve ambiguities between the presence of a glitch or an MBHB signal in each stretch, the second phase of the search evaluates the Bayes factor comparing the glitch and MBHB hypotheses.

\subsubsection{Second phase\label{search_second_phase}}
\noindent The first phase of our search aims to identify the presence of a signal in each stretch of data. In the second phase, we determine whether the signal is better described as a glitch or an MBHB. To this end, we perform a dedicated MBHB search to assess the presence of a GW signal in stretches where a signal was detected or where the estimated SNR is sufficiently high (N6, N23, N24,  N26, N27, N28, and N29). We then compute the Bayes factor comparing the MBHB and glitch hypotheses\footnote{Note that this analysis could alternatively be performed using the RJ-MCMC framework, which would allow for automatic computation of Bayes factors between glitches and MBHBs by leveraging the \texttt{Erebor} Global Fit infrastructure \cite{katz2024efficientgpuacceleratedmultisourceglobal}. However, incorporating this additional complexity is not necessary for the current analysis.}. For this, we employ both the stepping-stone method \cite{Maturana_Russel_2019} and thermodynamic integration \cite{Karnesis_2023}. 
This information is subsequently used to initialise the walkers for the parameter estimation phase. We then compare the Bayes factors between the glitch and MBHB hypotheses for each of these stretches. To this end, we use 30 temperatures and run the sampler with a fixed-dimensional parameter space. We fit only for the signal, assuming the noise to be at a fiducial level, following the strategy outlined in \cite{katz2024efficientgpuacceleratedmultisourceglobal}. This approach is robust given the high SNR of both glitches and MBHBs, and no ambiguous Bayes factor values were observed that would suggest the simultaneous presence of both components. In potential edge cases, the existing methodology in the \texttt{Erebor} Global Fit framework \cite{katz2024efficientgpuacceleratedmultisourceglobal} offers a viable and complementary strategy for MBHB detection. \\

\noindent We show in Table~\ref{tab:bayes_factors} three representative examples of the log Bayes factor for stretches N6, N24, and N29, where a signal (either a glitch or an MBHB) was identified. For brevity, we report only the results obtained using the stepping-stone method. In the Appendix \ref{bayes_factor} are reported the results obtained using the thermodynamic integration for completeness. The Bayes factors across all considered stretches were found to be highly decisive, strongly favouring one hypothesis over the other, either the presence of a glitch or that of an MBHB. For this reason, we present only a subset of representative results here, as the conclusions were clear-cut in all cases. Specifically, for stretches N6 and N24, the presence of a glitch is strongly favoured, while for stretch N29, the MBHB hypothesis is preferred.

\begin{table}[h]
    \centering
    \begin{tabular}{|lccc|}
        \hline
        & \textbf{N6} & \textbf{N24} & \textbf{N29} \\
        \hline
        $\log_{10} \mathcal{B}_{\mathrm{MBHB}/\mathrm{glitch}}$ 
        & $\sim -6.1\times 10^4$ 
        & $\sim -3.9\times 10^4$ 
        & $\sim 1.5\times 10^7$ \\
        
        $d\log_{10}  \mathcal{B}_{\mathrm{MBHB}/\mathrm{glitch}}$ 
        & $\sim 0.61$ 
        & $\sim 0.22$ 
        & $\sim 0.44$ \\
        \hline
    \end{tabular}
    \caption{Log Bayes factors comparing the MBHB and glitch models for datasets N6, N24, and N29 using the stepping-stone method. $d\log_{10}  \mathcal{B}_{\mathrm{MBHB}/\mathrm{glitch}}$ gives an estimate on the error on $\log_{10}  \mathcal{B}_{\mathrm{MBHB}/\mathrm{glitch}}$.}
    \label{tab:bayes_factors}
\end{table}

\noindent During the search phase, along with computing the Bayes factor, we also obtain a rough estimate of the parameters of the MBHB. However, doing parameter estimation for the MBHB using only a day of data highlights several key considerations regarding frequency domain analysis and the use of the discrete Fourier transform. One primary concern is spectral leakage, particularly evident when the pre-merger/inspiral phase of the signal is analyzed, which necessitates the use of windowing techniques to mitigate its effects \cite{davies2024premergerobservationcharacterizationmassive}. Indeed, when analyzing frequency-domain waveforms over short time segments, particular caution is required to ensure accuracy. However, in the context of this research, we did not window the data as it was not strictly necessary since we were not interested in characterising the pre-merger part of the signals\footnote{As illustrated in Appendix \ref{app:filter discussion}, filtering the data favours no correlation between adjacent frequency bins.}. These issues reflect a fundamental challenge in performing low-latency analyses of MBHB signals \cite{Cornish:2021smq}, but once again, we do not need to be so careful in the search phase, as the final characterisation of the signals uses the full data set. \\

\noindent To conclude on a more technical side, to be able to perform the analysis of the MBHB with BBHx but restricting it to a single day, we need to account for the time window considered when generating the MBHB template with BBHx. This is achieved by shifting the phase with respect to the starting time of our window, 
and by defining 
a start and an end time of the observation 
to cut the waveform, such that any frequency components of the signal that are emitted outside of this range are excluded.

\subsection{Parameter estimation phase\label{pe_phase}}
\noindent For the final parameter estimation phase we consider the full month of data. 
\noindent We include an MBHB template in our likelihood and consider the same prior as the one reported in Table~\ref{mbmb_table}, but restrict the prior on the time of coalescence between $\rm t_{\rm ref}-24  \rm s < t_{\rm ref} < t_{\rm ref}+24 \rm s$ according to the MBHB posteriors found during the search phase. We also consider a slightly restricted prior range for the glitch parameters based on the findings from the search phase, as shown in Table~\ref{prior_on_glitch_PE}. In particular, we restrict the times of injection and exclude combinations of $\beta_0$ and $A_0$ that do not reflect the glitch distribution found in LPF. If such glitches were present, they would have already been identified during the search phase, given the high SNR of those glitches. \\
We then set the minimum number of glitches for the reversible jump to match the number of high SNR glitches identified in the search phase, and we set the maximum number of glitches to be double the minimum to allow for glitches that could have been missed during the search. \\

\noindent We consider the same in-model and out-of-model moves -that were used during the $first$ search phase- for the glitches. The in-model moves get updated every 20 iterations up to a total of 50000 iterations. We also incorporate the covariance matrices computed from the three estimated glitches -during the search phase- to define an additional reversible jump move proposal (see section~\ref{ref:rj_move_glitch}) which proposes the birth or death of a glitch based on those identified during the search. 
To be on the safe side, we still allow $40\%$ of proposed glitch parameters to be drawn from the full prior range, ensuring that we do not miss any potential glitches. \\
\noindent Instead, for the in-model Gaussian move proposal of the MBHB we estimate the covariance matrix using samples from the MBHB parameter posteriors obtained during the search phase. \\ \noindent Finally, we initialise the MCMC walkers using samples found during the search phase, and we estimate the parameters of the noise, signal, and glitches simultaneously following a Gibbs sampling approach \cite{Karnesis_2023}. 

\begin{figure*}
\centering
\includegraphics[width=\textwidth]{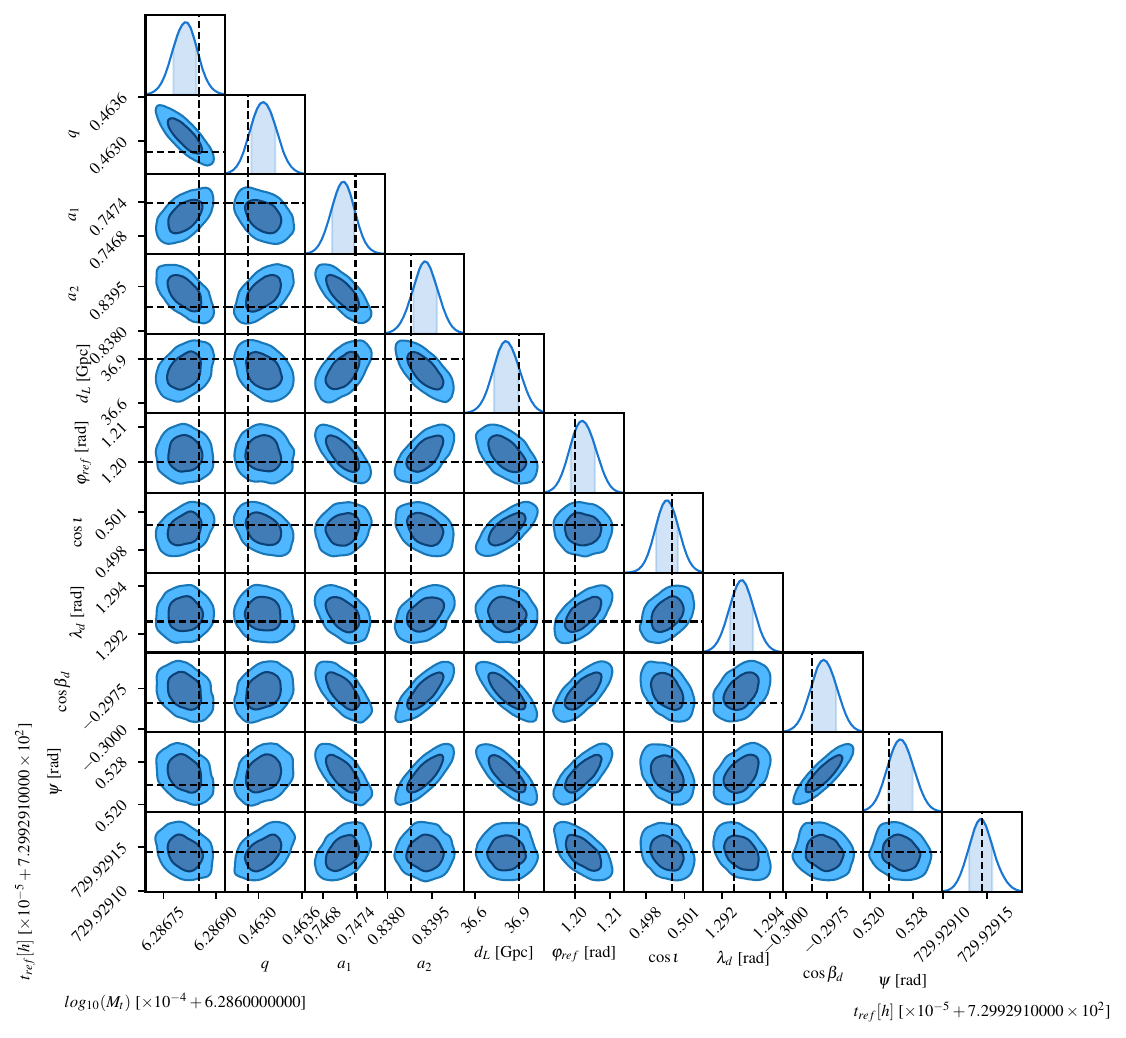}
\caption{Posteriors of the MBHB in the ``\textit{Light}-Spritz'' data set.\label{fig:corner_plot_mbhb_spritz}}
\end{figure*}

\begin{figure}[ht!]
\centering
\includegraphics[width=\columnwidth]{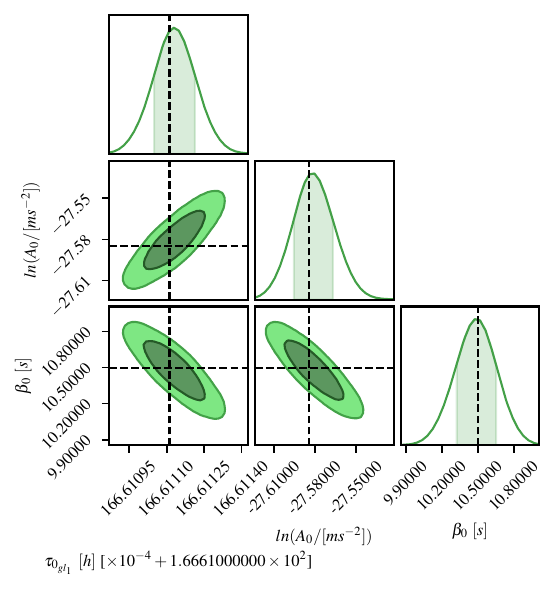}
\caption{Posteriors of the glitch injected around 166 hours in the ``\textit{Light}-Spritz'' data set.\label{fig:corner_plot_glitch1_spritz} }
\end{figure}

\begin{figure}[ht!]
\centering
\includegraphics[width=\columnwidth]{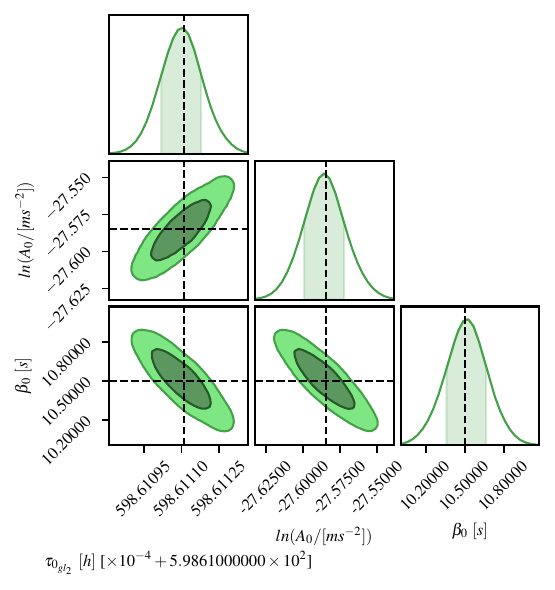}
\caption{Posteriors of the glitch injected around 598 hours in the ``\textit{Light}-Spritz'' data set.\label{fig:corner_plot_glitch2_spritz}}
\end{figure}

\begin{figure}[ht!]
\centering
\includegraphics[width=\columnwidth]{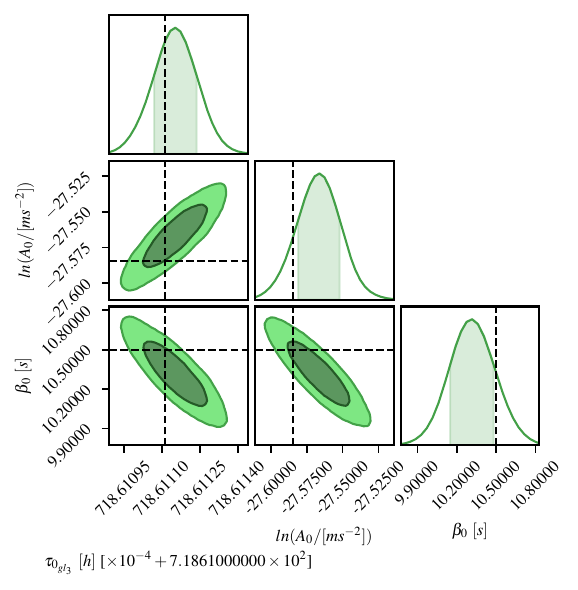}
\caption{Posteriors of the glitch injected around 718 hours in the ``\textit{Light}-Spritz'' data set. \label{fig:corner_plot_glitch3_spritz}}
\end{figure}

\begin{figure}[ht!]
\centering
\includegraphics[width=\columnwidth]{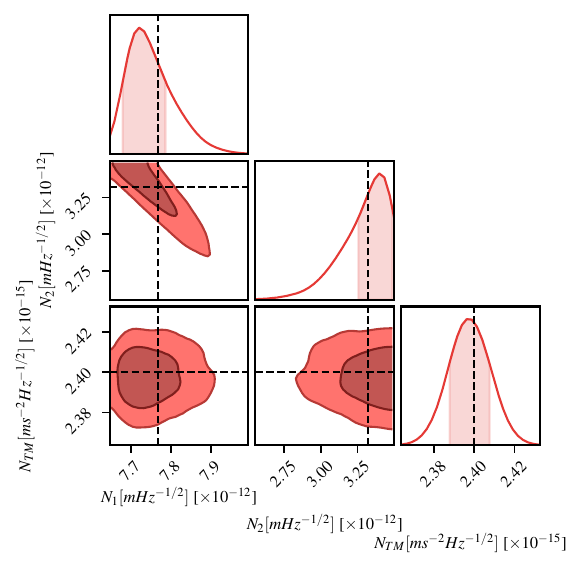}
\caption{Posteriors of noise parameters estimated in the ``\textit{Light}-Spritz'' data set.\label{fig:corner_plot_noise_spritz}}
\end{figure}

\noindent The posteriors for the MBHB are shown in Figs. \ref{fig:corner_plot_mbhb_spritz}, for the three glitches are shown in Figs.
\ref{fig:corner_plot_glitch1_spritz},\ref{fig:corner_plot_glitch2_spritz},\ref{fig:corner_plot_glitch3_spritz} and for the noise parameters in Fig. \ref{fig:corner_plot_noise_spritz}. The MBHB has an SNR of 1522 while the glitches have SNRs of around 300.  For all the posteriors, we only plot the last 500 samples of each walker, as we consider the previous samples, where we perform recursive updates of the in-model moves, as a burn-in period. 

\begin{table}[ht!]
\begin{centering}
\begin{tabular}{|c|c|c|}
    \hline
   Parameter & Lower & Upper \\ 
    &Bound  & Bound \\
    \hline
    $\tau_{0_{gl_1,gl_2,gl_3}}$ [s] & 432000  & 2627995 \\ 
    \hline
     $ln(A_0/[m s^{-2}])$ & -32 & -20 \\ 
    \hline
      $\beta_0$ [s] & 1 & 1e4\\  
      \hline
     $N_1$ [$\si{\meter\per\sqrt\hertz}$] & \num{7e-12}& \num{8e-12}\\
     \hline  
     $N_2$ [$\si{\meter\per\sqrt\hertz}$] & \num{2.5e-12}  & \num{3.5e-12} \\ 
     \hline  
     $N_{\rm TM}$  [$\si{\meter\per\second\squared\per\sqrt\hertz}$]& \num{2.2e-15}  & \num{2.7e-15}  \\
     \hline
\end{tabular}
\caption{Table of priors for the glitch and noise parameters. We consider a single broad prior on $\tau_{0_{gl_1,gl_2,gl_3}}$ that encompasses all three glitches injection times.} \label{prior_on_glitch_PE}
\end{centering}
\end{table}

\section{Conclusion and future perspective\label{sec:final}}
\noindent In this work, we have demonstrated a potential Global Fit approach for LISA to estimate MBHB parameters in the presence of noise artefacts by implementing a search pipeline to find and fit for glitches along with the noise properties of the instrument and the parameters of an MBHB, using realistic simulated LISA data. Our methodology provides a general framework for detecting and modelling glitches, ensuring that such instrumental artefacts do not compromise LISA's scientific output. \\  

\noindent We have developed and validated our pipeline in two steps. First, we generated synthetic data where multiple glitches of varying SNRs, including those closely spaced in time, were injected to test the effectiveness of our reversible jump approach in handling diverse glitch morphologies. Our injected glitches were randomly drawn from the LPF glitch distribution to provide a realistic test scenario. We established an approximate threshold for distinguishing glitches from instrumental noise and investigated the impact of low, medium and high SNR glitches on the parameter estimation of a fiducial MBHB with an SNR of 1152. We found that unfitted glitches with SNR $\gtrsim$ 60 introduce non-negligible biases in MBHB parameter recovery. For instance, for an unfitted glitch with SNR 21 we observed a slight shift in the parameter posterior relative to the injected values, but this shift was small compared to the statistical uncertainties, whereas for a glitch with SNR 72, the true values were excluded at high confidence from the posterior support for several parameters, such as the total mass or the sky location.  
For the extreme case of an SNR 1544 glitch, biases were very large, and the posterior was completely disjointed from the case where no glitch is present. Moreover, we noticed that the relative SNR between the glitch and the MBHB signal, as well as the glitch's temporal proximity to the merger, are critical factors in determining the extent of bias. However, these are not the sole contributors. For instance, when comparing the posterior distributions of MBHB systems with SNRs 1152 and 320, the glitch number $\#41$ in the LPF catalogue appears to cause a more pronounced bias in the recovery of parameters for the low-SNR system, hinting that also the relationship between the glitch frequency and the instananeous frequency of the MBHB plays a role. When fitting for the glitch(-es) (single glitch as well as multiple glitches), we corrected for these biases and estimated the parameters of the MBHB correctly, with almost the same accuracy when no glitch was present. This study stresses the importance of accounting for glitches to fully exploit the tremendous scientific potential of MBHB observations with LISA—whether for fundamental physics~\cite{Berti:2005ys,Berti:2016lat,Chamberlain:2017fjl,Barausse:2020rsu,Maggio:2020jml,Bhagwat:2021kwv,Corman:2021avn,Toubiana:2023cwr,Pitte:2024zbi}, astrophysics~\cite{Gair:2010bx,Sesana:2010wy,Klein:2015hvg,Dayal:2018gwg,Barausse:2020mdt,Toubiana:2021iuw,Chen:2022sae,Fang:2022cso,Spadaro:2024tve,Langen:2024ygz,Toubiana:2024bil}, or cosmology~\cite{Caprini:2016qxs,Tamanini:2016zlh,Mangiagli:2023ize}.\\

\noindent Second, we applied our search algorithm to a simplified version of the ``Spritz'' dataset, which was obtained by removing the Galactic binaries and gaps from the original dataset. Our results show that we can successfully implement a search and a parameter estimation pipeline to detect and model glitches present in the data while recovering both the instrumental noise and the MBHB parameters without bias. Additionally, we provided a detailed noise model for this dataset after identifying inconsistencies between the official LDC ``Spritz'' documentation~\cite{LDCS} and the actual procedure used to generate data for this mock dataset. This information will be valuable for future researchers analysing the same dataset. \\ 

\noindent A key aspect that warrants further investigation is the accuracy and efficiency of the GMM clustering approach used to build proposal distributions for glitches in the parameter estimation phase, particularly in determining the number of components. This is particularly important in view of including the search and parameter estimation for glitches, developed in this work, in the framework of the \texttt{Erebor} Global Fit \cite{katz2024efficientgpuacceleratedmultisourceglobal}, in which multiple MBHBs and other types of GW sources are present. In the current implementation, the number of clusters is fixed to match the maximum number of leaves (components) in the model, without attempting to infer the true number of underlying glitch populations. While this approach offers flexibility and has demonstrated to work very well in our framework, it may introduce inefficiencies when the number of GMM components significantly exceeds the actual number of distinct glitches. More specifically, in scenarios where the GMM overestimates the number of glitch clusters, the model may fragment individual glitch posteriors into sub-posteriors, which can hinder efficient sampling. As a result, MCMC chains may struggle to traverse the full posterior landscape, especially across different glitch modes, even if local acceptance rates remain high. We have begun exploring alternative strategies to dynamically estimate the appropriate number of clusters, potentially improving the exchange of samples across posterior modes and enhancing the overall sampling efficiency. This might play a more significant role when more GW sources or more complicated noise components, with respect to the ``\textit{Light}-Spritz'' data set, are considered.\\

\noindent Nonetheless, the glitch search, determination, and parameter estimation components developed in this work will form the foundation of the glitch handling module within \texttt{Erebor}. From \texttt{Erebor}, we will incorporate the already developed modules for MBHB searches and parameter estimation, as well as the search and parameter estimation for Galactic binaries, noise and foreground fitting. Over time, we expect to refine both the glitch and MBHB modules based on the insights gained here, particularly in trans-dimensional sampling strategies. We will also update the noise fitting by leveraging the more complete noise models reported in this work. \\

\noindent Future work will also extend this approach to account for instrumental non-stationarities, such as data gaps and the presence of the Galactic foreground, so that data with the more realistic properties of the ``Spritz'' data set can be handled. While our study serves as a guideline for glitch mitigation in LISA, it does not encompass all possible MBHB populations, and our reference glitch distribution is based on LPF data. However, the shapelet-based approach should readily adapt to different glitch morphologies, with respect to the signal exponential ones addressed in this paper, present in the data in particular by leveraging higher order shapelets, but its efficacy needs to be properly quantified. However, since LISA will use similar technology to LPF, at least for the GRS, the acceleration glitches are unlikely to differ substantially, so the results obtained here should represent the expected performance on actual LISA data. \\

\noindent \noindent To conclude, in this work, we have adopted simplifying assumptions in constructing the orthogonal TDI combinations $A$, $E$, and $T$. Specifically, we assumed equal and constant arm lengths, and uncorrelated noise across different channels. These assumptions simplify the likelihood by yielding a diagonal covariance matrix and lead to identical TDI transfer functions for the $A$ and $E$ channels. These assumptions are appropriate for our analysis of the ``Spritz'' data, which we restricted to utilize the $A$ and $E$ channels, where the impact of using the equal-arm approximation is less severe than for the null-channel $T$. Moreover, we excluded in the analysis the high-frequency region near the nulls of the TDI transfer functions, where $A$ and $E$ would also be heavily affected; this should not meaningfully affect our results, as all signals considered here lie at lower frequencies.
While such idealisations are appropriate for the scope of this study, we acknowledge that in the actual LISA mission, the arm lengths will be unequal and time-varying, and noise correlations between channels might not be negligible. This is especially true for the high-frequency regime, and ongoing studies are investigating how to properly model these effects and incorporate non-stationarities arising from orbital dynamics. However, these effects are more relevant for sources such as stellar-origin black hole binaries inspiraling in the LISA band at high frequencies. For instance, preliminary investigations in \cite{Hartwig_2022} suggest that alternative TDI combinations with fewer delays than the standard Michelson channels $X$, $Y$, and $Z$ could be preferable, as the nulls in the TDI transfer function occur at higher frequencies with respect to the Michelson channels. For the low-frequency regime, the main improvement by moving away from our simplifying assumptions lies in being able to exploit a null channel, like $T$, which might allow us to put stronger constraints on the noise during the search phase or to better characterise it during the parameter estimation phase. \\

\noindent All code used in this work is open-source and can be accessed at \href{the artefacts repository}{https://github.com/martinaAEI/artifacts}. 

\begin{acknowledgments}
\noindent The authors thank the following people (alphabetically ordered) for useful discussion in improving the work: Quentin Baghi, Jean-Baptiste Bayle, Ollie Burke, Eleonora Castelli, Raffi Enficiaud, Hector Hestelles, Niklas Houba, Nikos Karnesis, Lorenzo Pompili, Lorenzo Sala, Elise Sangër, Alessandro Santini, Lorenzo Speri, Alice Spadaro, Mauro Pieroni and Sebastian Völkel. M.M. and O.H. gratefully acknowledges the support of the German Space Agency, DLR. The work is supported by the Federal Ministry for Economic Affairs and Climate Action based on a resolution of the German Bundestag (Project Ref. No. FKZ 50 OQ 2301). A.T. is supported by MUR Young Researchers Grant No. SOE2024-0000125, ERC Starting Grant No.~945155--GWmining, Cariplo Foundation Grant No.~2021-0555, MUR PRIN Grant No.~2022-Z9X4XS, MUR Grant ``Progetto Dipartimenti di Eccellenza 2023-2027'' (BiCoQ), and the ICSC National Research Centre funded by NextGenerationEU. 
\end{acknowledgments}
\newpage

\appendix
\section{Glitch model}
\subsection{Derivation of shapelet model in time and frequency domains \label{shapelet_derivation}}

\noindent Shapelets can be defined in the time domain as follows:
\begin{equation}\label{eq:com_shap}
\psi_n(t) = 2 \frac{t}{n} e^{-\frac{t}{n}}L^1_{n-1}\bigg (2 \frac{t}{n}\bigg)\Theta(t),
\end{equation}
with $L_n^\alpha$ denoting a generalised Laguerre polynomial. A glitch perturbation is then modelled by a linear combination of shapelets:
\begin{equation}\label{eq:glitch_shap}
    g(t) = \sum_{j = 0}^{P} A_j \psi_{n_j} \bigg (\frac{t -\tau_j}{ \beta_j}\bigg),
\end{equation}
with $P+1$ being the number of shapelets. Substituting Eq.~\eqref{eq:com_shap} into Eq.~\eqref{eq:glitch_shap} we have:
\begin{align} \label{eq:g(t)_lag}
    g(t) &= \sum_{j = 0}^{P} 2 A_j  \frac{1}{n_j} \frac{t -\tau_j}{\beta_j}  
    e^{-\frac{1}{n_j} \frac{t -\tau_j}{\beta_i}}  \notag \\
    &\quad \times L^1_{n_j-1} \bigg (2 \frac{1}{n_j} \frac{t -\tau_j}{\beta_j} \bigg ) 
    \Theta \bigg (\frac{t -\tau_j}{ \beta_j} \bigg ),
\end{align}
Note that Eq.~\eqref{eq:g(t)_lag} refers to a glitch in acceleration, but we need to derive the corresponding fractional frequency shift, which means integrating $g(t)$ and dividing by the velocity of light $c$. \\

\noindent To perform the integral, we can consider the generating function:\begin{equation}\label{eq:generating function}
\sum_{n=0}^{\infty}  v^n L_n^\alpha(x) = \frac{1}{(1-v)^{\alpha +1}} e^{-\frac{x v}{1-v} }.
\end{equation}
To align it with Eq.~\eqref{eq:com_shap}, we adjust the index of the series as follows:
\begin{equation}\label{eq:generating function_new_ser}
\sum_{n=1}^{\infty} v^n L_{n-1}^{\alpha}(x)  = \frac{v}{(1-v)^{\alpha + 1}}e^\frac{- x v}{1-v}.
\end{equation}
and then define
\begin{align}
    &G(t,\tau, n, \beta, A,v) \nonumber \\
    &\hspace{0.5cm}= \frac{2 A}{c} \int_0^t \left( \frac{t'-\tau}{n \beta}\right) \exp\left[{-\frac{t'-\tau}{n\beta}}\right] \nonumber \\
    &\hspace{1cm} \times \frac{v}{(1-v)^{\alpha+1}} \exp\left[{-2\frac{t'-\tau}{n\beta}  \frac{v}{(1-v)}}\right] \nonumber \\
    &\hspace{2cm} \times \Theta\left(\frac{t'-\tau}{\beta}\right) \, {\rm d}t'\label{genfncint}
\end{align}
This expression enables us to obtain different shapelets orders by rewriting the relationship between the left and right components of Eq.~\eqref{genfncint} as follows:
\begin{align}
    &\frac{\Delta \nu_g}{\nu_0}(t,\vec{\tau}, \vec{n}, \vec{\beta}, \vec{A}) 
    \equiv \int_0^t \frac{g(t')}{c} \, dt' \nonumber \\
    &\hspace{1cm}= \sum_{i=0}^P \frac{1}{n_i!} \left[ \frac{{\rm d}^{n_i}}{{\rm d}v^{n_i}} G(t,\tau_i, n_i, \beta_i, A_i,v)\right]_{|v=0}.
\end{align}
Eq.~(\ref{genfncint}) can be explicitely written as:
{\small
\begin{align}
  &G(t,\tau, n, \beta, A,v)
    = \frac{2 v A}{c (1-v)^{\alpha + 1}} n \beta 
   \nonumber \\
   &\hspace{2cm} \times \int_{-\frac{\tau}{n \beta}}^{\frac{t - \tau}{n \beta}} 
    u e^{-u} e^{\frac{-2 u v}{1 - v}}   \Theta(n u) \, {\rm d}u \nonumber   \\
    &\hspace{1cm}= - \frac{ 2 v n A \beta (1 - v)^{- \alpha}}{c  (1 + v)^2} \times  \Theta \bigg(\frac{t-\tau}{\beta}\bigg) \notag \\
    &\times \bigg( -1 + v 
    + \frac{e^{\frac{(1 + v)(t - \tau)}{ n(-1 + v)\beta}} 
    ( t + t v + n \beta - n v\beta -  (1 + v)\tau )}{n\beta} 
    \bigg). \label{eq:general_shap}
\end{align}
} \\

\noindent In this way, the generating function is used to perform the integral. Equation \ref{eq:general_shap} needs to be computed for each value of $A_j, \tau_j$ and $\beta_j$ that is needed such that the sum over the shapelets can be performed. \\

\noindent We can now derive the Fourier transform of $\frac{\Delta v_g}{\nu_0}(t,\vec{\tau},\vec{n},\vec{\beta},\vec{A})$. This reads:
\begin{align}
    &\frac{\Delta \tilde{\nu}_g}{\nu_0}(f,\vec{\tau}, \vec{n}, \vec{\beta}, \vec{A}) = \\
    &\hspace{1cm} \sum_{j=0}^P \frac{1}{n_j!} \left[ \frac{{\rm d}^{n_j}}{{\rm d}v^{n_j}} \tilde{G}(f,\tau_j, n_j, \beta_j, A_j,v)\right]_{|v=0}.
\end{align}
Considering Eq. \ref{eq:general_shap}, we obtain:
\begin{align}
 \tilde{G}(f,\tau, n, & \beta, A,v) =   \nonumber \\
& \frac{i A n (1 - v)^{1 - \alpha}v\beta e^{-2 i f\pi\tau} }{c  f \pi (i (1 + v) + 2 f n \pi (-1 + v) \beta)^2},
\end{align}
and therefore,
\small{\begin{align}
 &\frac{\Delta \tilde{\nu}_g}{\nu_0}(f,\vec{\tau}, \vec{n}, \vec{\beta}, \vec{A}) = \noindent \\
&\sum_{j=0}^P \frac{1}{n_j!} \left[ \frac{{\rm d}^{n_j}}{{\rm d}v^{n_j}} \bigg( \frac{i A n (1 - v)^{1 - \alpha}v\beta e^{-2 i f\pi\tau} }{c  f \pi (i (1 + v) + 2 f n \pi (-1 + v) \beta)^2}\bigg)\right]_{|v=0}.
\end{align}}
\noindent The first order shapelet ($j = 0$) in the frequency domain is
\begin{equation}\label{eq:glitch_freq}
\frac{\Delta \tilde{\nu}_g}{\nu_0}(f,\tau_0, \beta_0, A_0) =
      \frac{i A_0 \beta_0 e^{2 i f \pi \tau_0}}{c f \pi 
         (i - 2 f \pi \beta_0)^2}.
\end{equation}

\subsection{Glitch model in  time and frequency domain and respective TDI transfer functions\label{glitch_tdi}}
\noindent The effect of a typical LPF glitch is evaluated starting from a model for its differential acceleration, $\Delta g(t)$, \cite{Sala:2023hpr,Muratore:2021rwq}:
\begin{equation}
  h_g(t) = \frac{\Delta v   }{\tau^2}t{'} e^{-\frac{t{'}}{\tau}}\Theta(t'), \quad\quad t{'} = t - t_0
\end{equation}
where $t_0$ is the occurrence time, $\Theta(t)$ is the Heaviside step function, $\Delta v$ is the total transferred impulse per unit mass, and the $\tau$ parameter determines the duration of the event. We can also express the fractional frequency shift from the glitch as: 
\begin{equation}
 \frac{ \Delta \nu_g(t)}{\nu_0} = \frac{\Delta v   }{c} \Bigg{(}1 - \frac{e^\frac{-t + T}{\tau}(t - T + \tau)}{\tau}\Bigg{)}\Theta(t - t_0).
\end{equation}
From a frequency-domain perspective, $\Delta v$ is also the zero-frequency limit of the Fourier transform of the $h_g(t)$ template. This is relevant, since it implies that it has a strong low-frequency component. the Fourier transforms of the template is:
\begin{equation}
\frac{\Delta \nu_g(f) }{\nu_0}=  -  \frac{\Delta v e^{-i 2 \pi f  t_0}}{c i 2 \pi f \left(-i  +  2 \pi f \tau \right)^2}\label{eq:glitch_freq},
\end{equation}
where we need to divide by $c \omega$ to express it as fractional frequency shift. The expression is similar to the shapelet when $n_0=1$ (Eq. \ref{Eq.shap_freq}) with $\Delta v = 2 A \beta$ and $t_0 = \tau$ and $\tau = \beta$. \\

\noindent Considering now Eq. \ref{eq:tdi1-definition} in section \ref{sec:tdi_tf}, we can model the TDI response to a glitch as a product between the glitch expression in the frequency domain (Eq. \ref{eq:glitch_freq}) and the transfer function $\mathcal{T}_{X,Y,Z}^{g}(f)$ which takes into account TDI (see section \ref{sec:tdi_tf}). We provide below the expressions for the first-generation TDI transfer functions, noting $\mathcal{F}{D_{pq}} = e^{-i (2 \pi f) L}$: \\

\noindent In case of a glitch happening on $TM_{21}$ we have:
\begin{subequations}
\begin{align}
\mathcal{T}_{X_1}^{g}(f) & = -2e^{-3 i L (2 \pi f)} (-1 + e^{2 i L (2 \pi f)})  \\
\mathcal{T}_{Y_1}^{g}(f) & = 1 - e^{-4 i L (2 \pi f)} \\
\mathcal{T}_{Z_1}^{g}(f) & = 0
\end{align}
\end{subequations}
In case the same glitch is happening on $TM_{13}$
\begin{subequations}
\begin{align}
\mathcal{T}_{X_1}^{g}(f) & = 1 - e^{-4 i L (2 \pi f)}  \\
\mathcal{T}_{Y_1}^{g}(f) & = 0 \\
\mathcal{T}_{Z_1}^{g}(f) & =  -2e^{-3 i L (2 \pi f)} (-1 + e^{2 i L (2 \pi f)}) 
\end{align}
\end{subequations}
or on $TM_{31}$
\begin{subequations}
\begin{align}
\mathcal{T}_{X_1}^{g}(f) & = 4 i e^{-2 i L (2 \pi f)} \sin(L (2 \pi f)) \\
\mathcal{T}_{Y_1}^{g}(f) & = 0 \\
\mathcal{T}_{Z_1}^{g}(f) & =  -1 + e^{-4 i L (2 \pi f)}  
\end{align}
\end{subequations}
or on $TM_{23}$
\begin{subequations}
\begin{align}
\mathcal{T}_{X_1}^{g}(f) & =0  \\
\mathcal{T}_{Y_1}^{g}(f) & = -1 + e^{-4 i L (2 \pi f)} \\
\mathcal{T}_{Z_1}^{g}(f) & = 4 i e^{-2 i L (2 \pi f)} \sin(L (2 \pi f))     
\end{align}
\end{subequations}
and finally in case of $TM_{23}$:
\begin{subequations}
\begin{align}
\mathcal{T}_{X_1}^{g}(f) & =0  \\
\mathcal{T}_{Y_1}^{g}(f) & =  -2e^{-3 i L (2 \pi f)} (-1 + e^{2 i L (2 \pi f)})  \\
\mathcal{T}_{Z_1}^{g}(f) & = 1 - e^{-4 i L (2 \pi f)}    
\end{align}
\end{subequations}
For the second generation TDI, a similar computation can be done using Eq. \ref{eq:tdi2-definition} in section \ref{sec:tdi_tf}.

\section{Posteriors of SNR 1152 MBHB with high SNR glitch of 1544 at different injection times with respect to its merger time \label{sec:mbhb_glitch}}

\begin{figure*}[ht!]
\centering
\includegraphics[width=\textwidth]{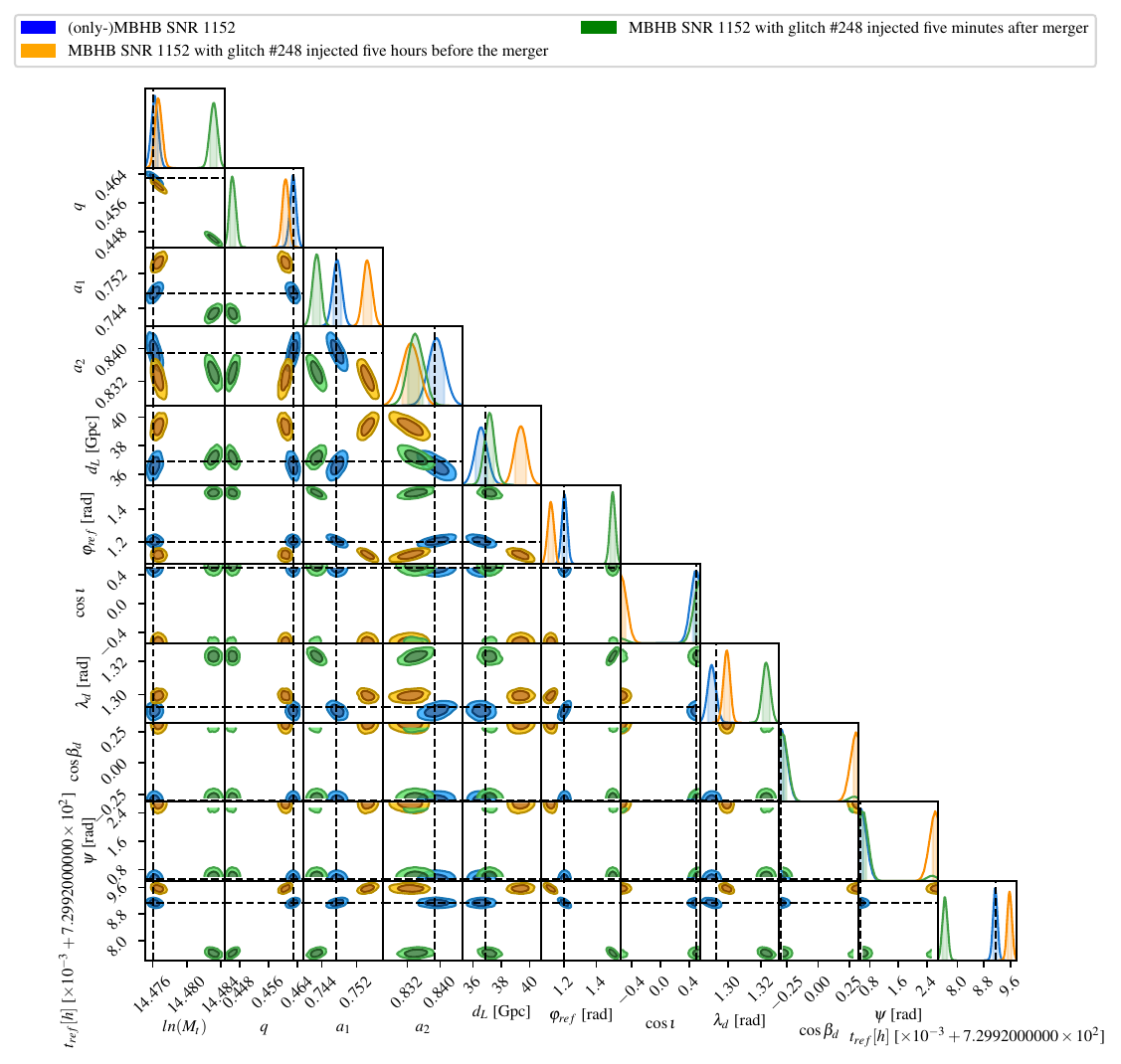}
\caption{Posteriors of MBHB 1152 SNR along with posterior of the same MBHB with a 1544 SNR glitch ($\#$ 248) injected five minutes after the merger time and injected five hours before the merger time\label{posterior_mbh_before_after}}
\end{figure*}

\noindent The impact of a glitch in the recovery of an MBHB is maximised when the glitch occurs closer to the merger; this, in practice, means that the same type of glitch would not have as great an impact if it happened at a different time. To show this, we also report below the impact on parameter estimation of the fiducial MBHB with SNR 1152 and redshift 4 -- considered also in the analysis of the ``\textit{Light}-Spritz'' data set -- with the 1544 SNR glitch injected at different times with respect to maximum overlap computed in section \ref{sec:estimation_part}: five hours before the merger and five minutes after the merger (Fig.~\ref{posterior_mbh_before_after}). In all cases, noisy data have been used, which introduce some random statistical fluctuations that are visible in the posteriors. \\

\noindent Looking at Fig.~\ref{posterior_mbh_before_after}, where the glitch has been injected five hours before the merger time, we can see that not fitting for the glitch causes less bias with respect to the glitch occurring five minutes after the merger. We expect that glitches happening earlier would have an even smaller effect. If the glitch occurs closer to or at the merger, not fitting for it clearly biases the parameters of the MBHB, such as the masses, spins or sky positions. This behaviour stems from the fact that the merger is the loudest part of the signal, and drives the parameter estimation of the MBHB, so glitches that corrupt that part of the signal have a stronger effect.

\newpage
\section{Posteriors of SNR 320 MBHB with glitches $N41$ and $N3$ injected at their maximum overlap \label{sec:mbhb_glitch_second_case}}
\noindent In Fig. \ref{fig:corner_plot_mbhb_glitches} are reported the posteriors of an SNR 320 MBHB when the two glitches corresponding to glitch $\#41$ and $\#3$ of the LPF catalogue are injected. The parameters of the MBHB are reported in Table \ref{mbmb_table_second_case}. The priors are the same as Table \ref{mbmb_table}, but the prior on the total mass $M_t$ has been converted to $Log_e$, i.e $(ln)$.\\

\begin{figure}[ht!]
\centering
\includegraphics[width=\columnwidth]{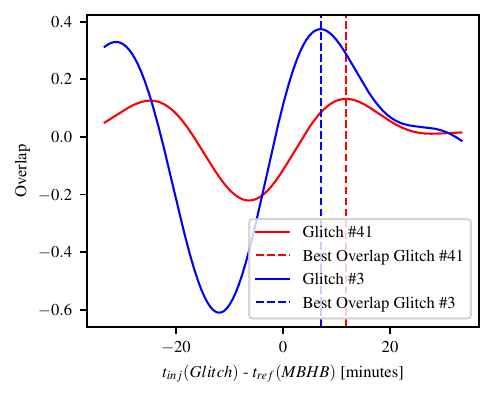}
\caption{Overlap between the MBHB and glitch signals as a function of the glitch injection time relative to the merger, where the decay time and amplitude of the glitches are kept fixed. The red and blue solid curves are for the two glitches with SNRs of 21 and 72. The GW signal is fixed to that of the parameters in Table \ref{mbmb_table_second_case}. Dashed vertical lines with matching colours denote the times that maximise the Overlap.\label{overlap_second_case}}
\end{figure}

\begin{figure*}[ht!]
\centering
\includegraphics[width=\columnwidth]{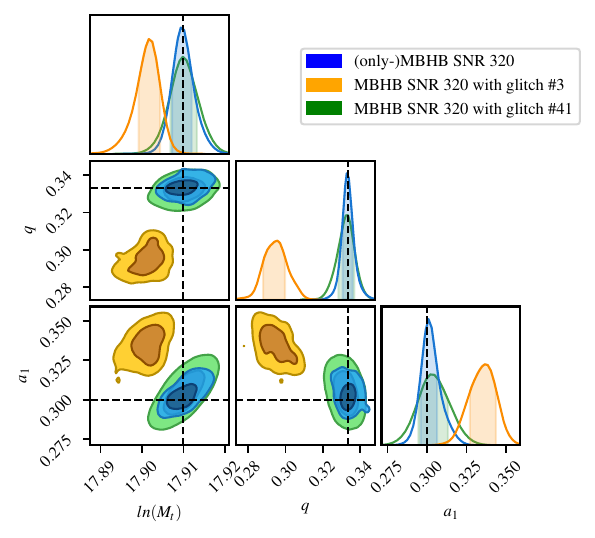}
\includegraphics[width=\columnwidth]
{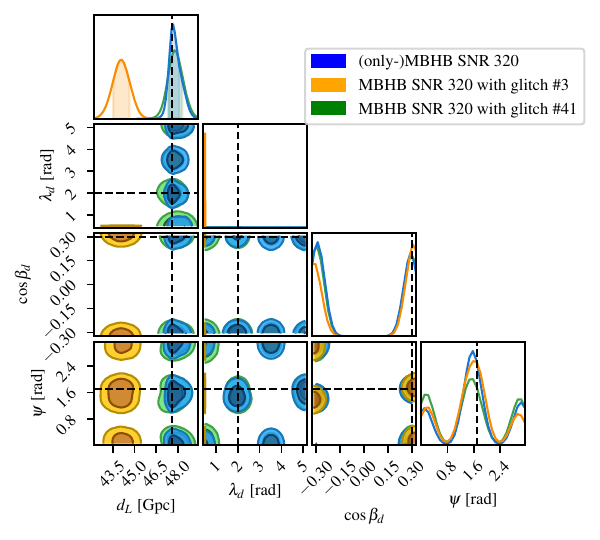}
\caption{Posteriors of a subsets of intrinsic parameters (left plot) and a subsets of extrinsic parameter (right plot) for the parameters of an MBHB with SNR 320 obtained from data containing only the MBHB, compared to those obtained from data containing a glitch of SNR 21 and of SNR 72, when the latter are not fitted.\label{fig:corner_plot_mbhb_glitches_Alice} 
}
\end{figure*}

\noindent The glitches were injected at their maximum Overlap accordingly to Eq. \ref{eq:maximum_overlap}, that is, 0.19 hours for glitch $\#41$ and 0.11 hours for glitch $\#3$, as can be seen from the Figure \ref{overlap_second_case}. In Fig. \ref{overlap_second_case}, we show the two glitches injected at the time that maximises the Overlap between the glitch and the fiducial MBHB in the TDI channel $A$. It is possible to observe that sky parameters have multimodality. Indeed, there are up to eight modes in the sky that are exactly degenerate for a fixed LISA configuration and non-evolving sources. However, this degeneracy is broken by LISA's motion and changes in frequency. High-mass and short-duration sources, like this case, are more likely to exhibit such degeneracies. The degeneracy primarily involves sky location and polarization  ($\beta_d$, $\psi$, $\lambda_d$ and $\cos \iota$)~\cite{Marsat_2021, PhysRevD.105.044055}, as can also be seen in the posteriors plot showing the extrinsic parameters (right plot Fig. \ref{fig:corner_plot_mbhb_glitches_Alice}) for the case of (only-)MBHB posteriors or when a low SNR 21 glitch is injected by not fitted. The left plot in Fig. \ref{fig:corner_plot_mbhb_glitches_Alice} shows the posteriors for the intrinsic parameters.
\begin{table}
\begin{centering}
\begin{tabular}{|c|c|}
    \hline
   Parameter & True Value\\ 
    \hline
    $ln(M_t)$ & 17.91  \\ 
    \hline
    $q$ & 0.33  \\ 
    \hline
      $a_1$ &0.3  \\  
      \hline
     $a_2$ &0.4  \\  
     \hline
      $d_L [Gpc]$ & 47.6  \\  
 \hline
      $\varphi_{ref}$ $[rad]$ & 1.2\\  
\hline
      $\cos \iota$ & 0.29  \\  
    \hline
      $\lambda$ $[rad]$ & 2.0  \\  
\hline
      $\sin \beta$ &0.82  \\  
 \hline
   $\psi$ $[rad]$& 1.6  \\  
  \hline
   $t_{ref} [h]$ & 30 \\  
   \hline
\end{tabular}
\end{centering} \caption{Massive Black Hole Binary true values. Values of the masses are given in the detector frame\label{mbmb_table_second_case}. We express the masses as dimensionless multiples of the solar mass $\rm M_{\odot}$.}
\end{table}

\newpage
\section{Discussion on posteriors for noise estimated during the search phase\label{search_results}}
\begin{figure}[ht!]
\centering
\includegraphics[width=\columnwidth]{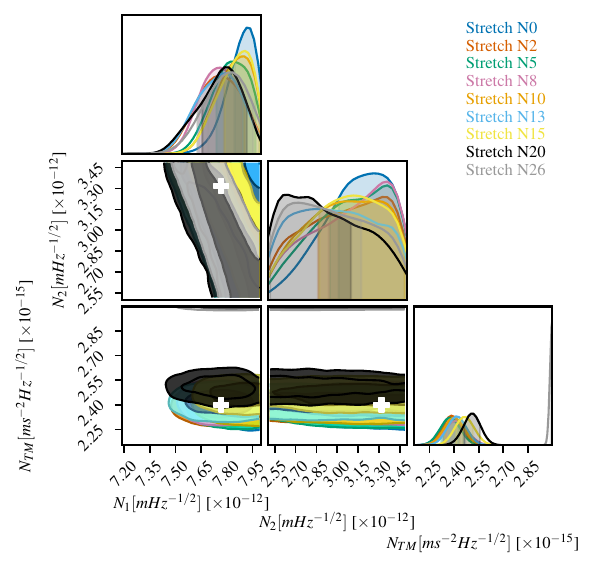}
\caption{Estimation of amplitude for the components of the test mass acceleration noise, test mass interferometer, inter-spacecraft interferometer and reference interferometer.\label{PE_noise_streches}}
\end{figure}

\begin{figure}[ht!]
\centering
\includegraphics[width=\columnwidth]{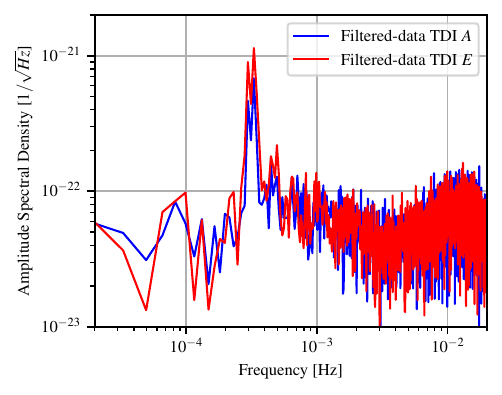}
\caption{Amplitude spectral density of the TDI $A$ and $E$ estimated during the search phase for the stretch N26, where the inspiral-phase of the MBHB is present.\label{searches_insp_n26}}
\end{figure}

\noindent Fig.~\ref{PE_noise_streches} shows the posterior distributions for the three noise parameters ($N_1$, $N_2$, and $N_{\mathrm{TM}}$), estimated during the search for segments N0, N2, N5, N8, N10, N13, N15, N20, and N26, along with the true values reported in section~\ref{sec:spritz-noise}. The TM acceleration noise is not well characterised, showing a bias away from the true value. This bias increases for data chunks that include the late inspiral phase of the MBHB (segments N26–N28), as this occurs in the frequency range around $10^{-3}$ Hz where TM noise dominates, as also visible in Fig.~\ref{searches_insp_n26}.\\
The $N_2$ component spans the entire prior range, indicating poor constraint. In contrast, $N_1$ is better estimated, as it dominates at higher frequencies. Even with just about a day of data, its value can be meaningfully constrained. On the other hand, the TMI noise remains poorly characterized, as its contribution lies in a frequency range where it is overshadowed by the stronger OMS noise, as discussed in section~\ref{sec:spritz-noise}, or in other words the OMS noise that enters in $N_2$ is much smaller than the OMS noise that contributes to $N_1$. Longer data segments are needed to improve the inference of TMI noise, as demonstrated in section~\ref{sec:estimation_part}, where the fit is performed on the full ``\textit{Light}-Spritz'' dataset. Nonetheless, this initial estimate is useful as an indicator of a reasonable prior range to be used during the parameter estimation phase, where the full data set is analysed. 

\section{Orthogonality of fast-Fourier transform of ``Spritz''-like data.\label{app:filter discussion}}

\noindent In \cref{sec:spritz-intro}, we propose to low-pass filter the ``\textit{Light}-Spritz'' data before taking the FFT (Fast Fourier Transform) to better approximate a diagonal likelihood. To verify this approach, we produce a toy-model with simplified ``Spritz''-like noise in the time domain and investigate the covariance matrix of the FFT. We use the same sampling rate as in ``Spritz'', $f_s = \SI{0.2}{\hertz}$, and produce  500 realisations of a noise data stream of 4000 samples. The noise in this toy-model is the sum of the two components,
\begin{equation}
    n_{\mathrm{toy}} = n_{\mathrm{TM}} + n_{\mathrm{OMS}},
\end{equation}
where the noise PSDs of $n_{\mathrm{TM}}$ and $n_{\mathrm{OMS}}$ are given by $C_{\mathrm{acc}}^{\mathrm{ffd}}S_{\mathrm{TM}}$ and $C_{\mathrm{disp}}^{\mathrm{ffd}}S_{\mathrm{OMS}}^{\mathrm{ISI}}$, as described in \cref{sec:spritz-noise}. \\

\noindent We approximate these noise shapes by adding white noise time series, shaped using repeated applications of \texttt{cumsum} and \texttt{gradient} operations. We then inject this noise into the variable $\eta_{12}$ of the 2nd generation TDI Michelson combination (see Eq.~\eqref{eq:tdi2-definition}), evaluated with equal arms and computed using the open source software pytdi~\cite{staab_2023_8429119}. Expressed in terms of delay operators, this corresponds to
\begin{equation}
    n_{\mathrm{toy}}^{\mathrm{TDI}} = (1 - D^2)(1 - D^4) n_{\mathrm{toy}}.
\end{equation}
We then consider three scenarios:
\begin{enumerate}
    \item Directly compute the FFT without any pre-processing
    \item Compute the FFT after applying a window function (\texttt{hann} window)
    \item Compute the FFT after applying a low-pass filter, as described in \cref{sec:spritz-intro}.
\end{enumerate}
After computing the FFT series, we estimate the covariance matrix by computing
\begin{equation}
    C_{jk} = E[\mathrm{FFT}_j\mathrm{FFT}_k^*],
\end{equation}
where $\mathrm{FFT}_j$ is the j'th frequency bin of the FFT of $n_{\mathrm{toy}}^{\mathrm{TDI}}$, considering only positive frequencies. The expectation value is approximated by averaging over our 500 realisations. 
\begin{figure}[ht!]
\centering
\includegraphics[width=\columnwidth]{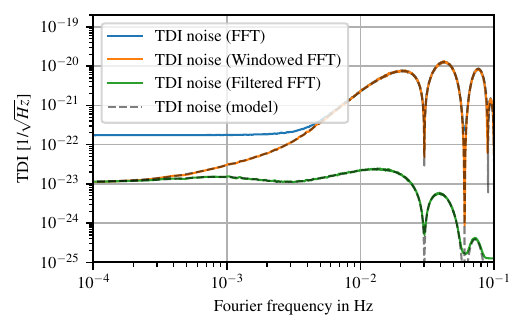}
\caption{Diagonal of FFT covariance matrix, scaled to Amplitude Spectral Density for a toy model equal and constant arm-length TDI $X_2$. \label{fig:fft-psd-estimation}}
\end{figure}
The square root of the diagonal elements of this matrix are displayed in Fig. \ref{fig:fft-psd-estimation}, rescaled by $\sqrt{\frac{2}{f_s \sum{w[j]^2}}}$ to approximate an ASD~\cite{heinzel-spectral-est}. Here, $w[j]$ are the window function coefficients applied to the j'th sample. \\

\noindent The direct FFT is strongly affected by spectral leakage and deviates from the expected ASD at low frequencies, whereas the windowed FFT and filtered FFT approximate their respective model well. Note that the low-pass filter is accounted for in the modelled noise shape for the latter.\\

\noindent To study correlations between frequency bins with the three methods, we normalise $C_{jk}$ by the values on the diagonal and plot
\begin{equation}
    \tilde C_{jk} = |C_{jk} / \sqrt{C_{jj}C_{kk}}|.
\end{equation}
The results are displayed in Fig. \ref{fig:fft-covariances}. For all three scenarios, we display the matrix for the full frequency range on the left as well as a zoomed-in version of just the first 100 frequency bins on the right. The top row of Fig. \ref{fig:fft-covariances} is the direct FFT. It shows strong long-range correlations around the frequencies corresponding to the zeros of the TDI transfer function as well as at DC, but otherwise a narrow diagonal. The middle row is the windowed FFT. The window strongly suppresses the long-range correlations, but introduces strong correlations in a narrow band along the diagonal. Finally, the filtered FFT shows even stronger correlations at high frequencies, but is perfectly diagonal in the band up to \SI{20}{\milli\hertz} used for our analysis. This validates the use of a diagonal likelihood model in section \ref{sec:spritz-intro}.

\begin{figure*}
    \centering
    \includegraphics[width=0.9\textwidth]{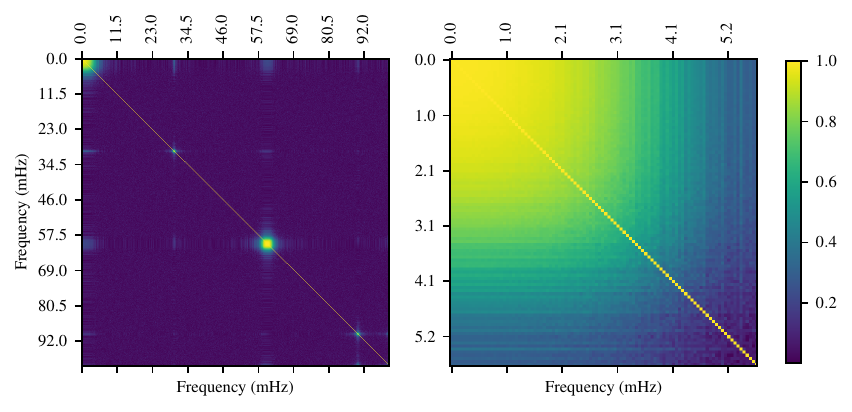}
    \label{fig:fft-covariances}
\vspace{1ex}
        \includegraphics[width=0.9\textwidth]{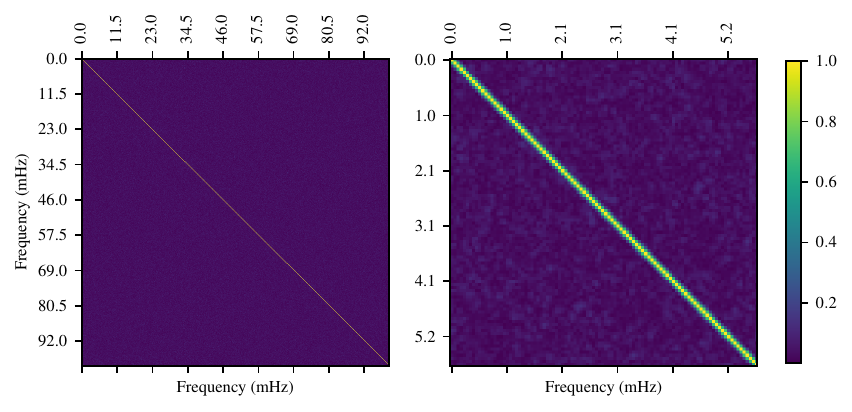}
\vspace{1ex}
        \includegraphics[width=0.9\textwidth]{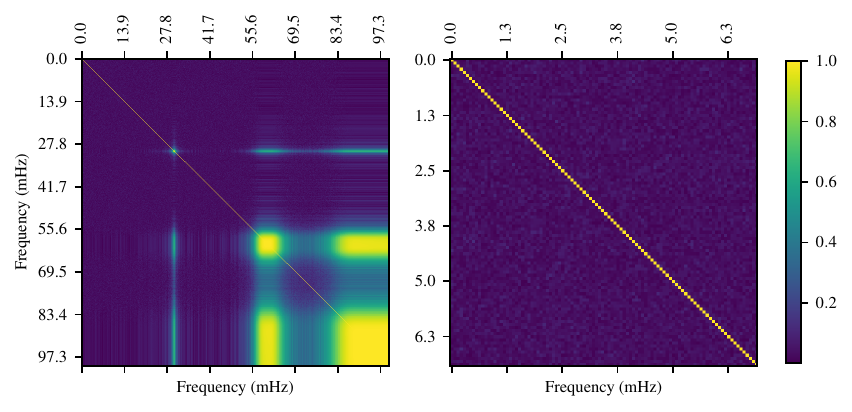}
    \caption{Numerically estimated covariance matrix of FFT. Top: no pre-processing, middle: window function, bottom: low-pass filter}
    \label{fig:fft-covariances}
\end{figure*}

\section{Bayes factor computation\label{bayes_factor}}
\begin{figure*}
\centering
\includegraphics[width=\columnwidth]{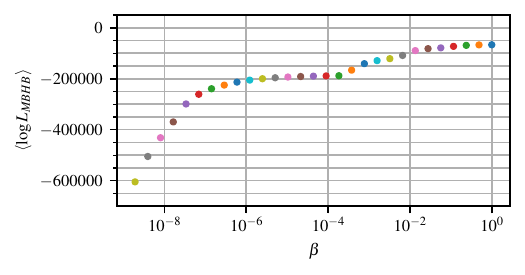}
  \includegraphics[width=\columnwidth]{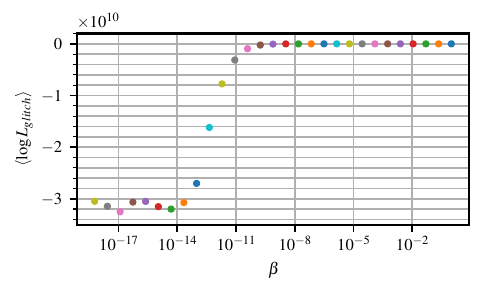}
\vspace{1ex}
\includegraphics[width=\columnwidth]{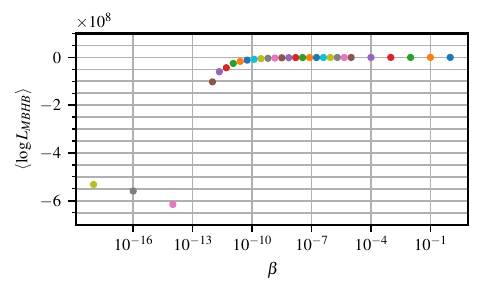}
  \includegraphics[width=\columnwidth]{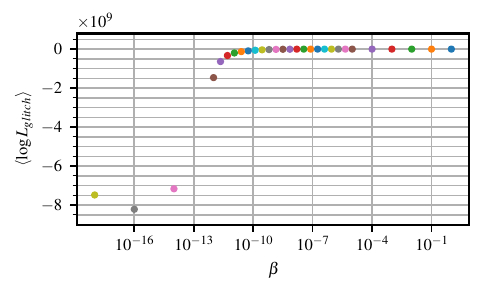}
  \vspace{1ex}
\includegraphics[width=\columnwidth]{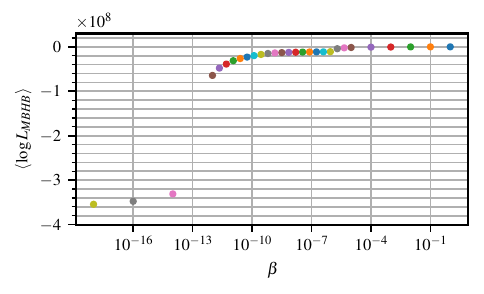}
  \includegraphics[width=\columnwidth]{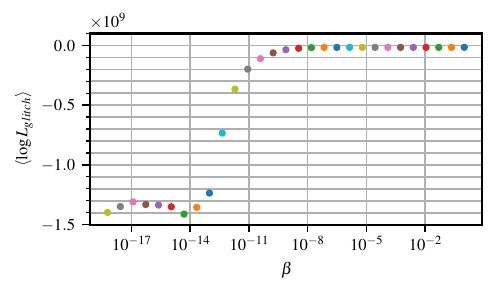}
  \caption{ The y-axis show the average log-likelihood (Log $L$) at each temperature for the MBHB (glitch) model, and the x-axis shows the inverse of the temperature. For the top row: $\log_{10} \mathcal{B}_{\mathrm{MBHB}/\mathrm{glitch}} \sim \num{-5.57e4}$ with $d\log_{10} \mathcal{B}_{\mathrm{MBHB}/\mathrm{glitch}} \sim 11843.7$ for stretch N6; for the middle row: 
    $\log_{10} \mathcal{B}_{\mathrm{MBHB}/\mathrm{glitch}} \sim \num{-2.93e4}$ with $d\log_{10} \mathcal{B}_{\mathrm{MBHB}/\mathrm{glitch}}  \sim 55796.5$ for stretch N24; and for the bottom row: $\log_{10} \mathcal{B}_{\mathrm{MBHB}/\mathrm{glitch}} \sim \num{1.56e7}$ with $d\log_{10} \mathcal{B}_{\mathrm{MBHB}/\mathrm{glitch}} \sim 28048.2$ for stretch N29. $d\log_{10}  \mathcal{B}_{\mathrm{MBHB}/\mathrm{glitch}}$ gives an estimate on the error on $\log_{10}  \mathcal{B}_{\mathrm{MBHB}/\mathrm{glitch}}$. Note that we have omitted the base 10 in the plots.\label{fig:bayes_factor}}
\end{figure*}

\noindent In Fig. \ref{fig:bayes_factor}, we report three examples of the stretches N6, N24 and N29 as an illustration of the methodology to compute the Bayes factor with thermodynamic integration. We use 70 temperatures to perform the thermodynamic integration method, where $\beta_k = \frac{1}{T_k}$ and $T_k$ represents the temperatures. As expected \cite{Karnesis_2023}, the stepping-stone algorithm produces more accurate estimates of the marginal likelihood than methods that use thermodynamic integration \cite{xie2011improving}; this is why in the computation shown here, the errors on the Bayes factor are higher. The y-axis shows the average log-likelihood (Log $L$) at each temperature for the MBHB (glitch) model, and the x-axis shows the inverse of the temperature. This curve is integrated to calculate the log evidence, log$_{10} Z$. The difference between the log evidence of the MBHB model versus the glitch model gives the log Bayes factor, $\rm \log_{10} \mathcal{B}_{\mathrm{MBHB}/\mathrm{glitch}}$.

\clearpage
\bibliography{MBHB_glitch_bibliography}

\end{document}